\documentclass[preprint,pra,showpacs,showkeys,floatfix]{revtex4}

\usepackage{graphicx}

\begin{document}

\title{WavePacket: A Matlab package for numerical quantum dynamics.\\
I: Closed quantum systems and discrete variable representations} 

\author{Burkhard Schmidt}
\email{burkhard.schmidt@fu-berlin.de}
\affiliation{
Institut f\"{u}r Mathematik, Freie Universit\"{a}t Berlin\\Arnimallee 6, D-14195 Berlin, Germany}

\author{Ulf Lorenz}
\email{ulf@wavepacket.org}
\affiliation{
Institut f\"{u}r Chemie, Universit\"{a}t Potsdam\\Karl-Liebknecht-Stra\ss{}e 24-25, D-14476 Potsdam, Germany}

\date{\today}

\begin{abstract}
WavePacket is an open-source program package for the numerical simulation of quantum-mechanical dynamics. 
It can be used to solve time-independent or time-dependent linear Schr\"{o}dinger and Liouville-von Neumann-equations in one or more dimensions. 
Also coupled equations can be treated, which allows to simulate molecular quantum dynamics beyond the Born-Oppenheimer approximation.
Optionally accounting for the interaction with external electric fields within the semiclassical dipole approximation, WavePacket can be used to simulate experiments involving tailored light pulses in photo-induced physics or chemistry.
The graphical capabilities allow visualization of quantum dynamics 'on the fly', including Wigner phase space representations. 
Being easy to use and highly versatile, WavePacket is well suited for the teaching of quantum mechanics as well as for research projects in atomic, molecular and optical physics or in physical or theoretical chemistry.
The present Part I deals with the description of closed quantum systems in terms of Schr\"{o}dinger equations. The emphasis is on discrete variable representations for spatial discretization as well as various techniques for temporal discretization.
The upcoming Part II will focus on open quantum systems and dimension reduction; it also describes the codes for optimal control of quantum dynamics.
The present work introduces the MATLAB version of WavePacket 5.2.1 which is hosted at the Sourceforge platform, where extensive Wiki-documentation as well as worked-out demonstration examples can be found. 
\end{abstract}

\maketitle

{\bf PROGRAM SUMMARY}

\begin{small}
\noindent
{\em Program Title: \textsc{WavePacket}}                            \\
{\em Licensing provisions: GPLv3}                                   \\
{\em Programming language: \textsc{Matlab}}                         \\

{\em Supplementary material: n.~a.}                                 \\
{\em Journal reference of previous version:  n.~a.}                 \\
{\em Does the new version supersede the previous version?: n.~a.}   \\
{\em Reasons for the new version: n.~a.}                            \\
{\em Summary of revisions: n.~a.}*                                  \\

{\em Nature of problem:}\\
  Schr\"{o}dinger's equations are of fundamental importance in non-relativistic quantum mechanics of distinguishable particles. 
	The solutions of the time-independent equation (TISE) are wavefunctions in coordinate space, the absolute squares of which are usually interpreted as probability density. 
	The time-dependent equation (TDSE) describes the dynamics of a quantum system evolving in time. 
	It plays a crucial role for the simulation, understanding, and prediction of modern experiments in atomic, molecular and optical physics where systems are driven by temporally shaped external fields.\\
{\em Solution method:}\\
   All numerical methods compiled in WavePacket are based on discrete variable representations. 
	Currently implemented are Gauss-Hermite, Gauss-Legendre and FFT-based schemes.
	The TISE is solved either by direct diagonalization or by propagation in imaginary time.
	For the TDSE there is a choice between second order differencing, operator splitting and Chebychev polynomial methods. \\
{\em Additional comments including Restrictions and Unusual features:}\\
  The \textsc{WavePacket} program package is rather easy and intuitive to use, providing visualization of quantum dynamics 'on the fly'. 
	It is mainly intended for low-dimensional systems, typically not exceeding three to five degrees of freedom.
	Detailed user guides and reference manuals are available through numerous Wiki pages hosted at the \textsc{SourceForge} platform where also a large number of well documented demonstration examples can be found.
\end{small}

\section{Introduction}

With the advent of time-dependent experimental techniques, most notably with the progress in generating short, intense laser pulses in the 1980s and 1990s, there has been a strongly increased interest in studying quantum mechanics also from the time-dependent point of view \cite{May:00a,Schleich:01a,Tannor:04a}.
Experiments using tailored laser fields have often been going hand in hand with quantum dynamical simulations, leading to enormous progress in the fields of atomic and molecular physics \cite{Grossmann2008}, femtochemistry \cite{Zewail2000} and even femtobiology \cite{Sundstrom2008}. 
More recently, concepts developed in these fields have been transferred to the study of quantum information and quantum computing \cite{Vivie:07a,Kais2014}.

Despite the obvious importance of quantum dynamics simulations, general and freely available simulation software is still rather scarce; one notable exception being the MCTDH program package which specializes in weakly coupled, high-dimensional systems \cite{Beck2000}.
In this work we introduce our general-purpose \textsc{Matlab} software package \textsc{WavePacket} (version 5.2.1) for numerical quantum dynamics, giving a step-by-step explanation of how typical simulation tasks can be carried out.

The \textsc{WavePacket} software package uses a straightforward approach by expanding all wave functions and relevant operators in finite basis representations (FBRs) and/or associated discrete variable representations (DVRs) \cite{Light:85a,Light:00a}.
The time-independent Schr\"odinger equation is then trivially cast into an eigenvalue problem whereas the time-dependent Schr\"odinger equation is solved by a variety of propagation methods, all of which make use of fast transformations between DVRs and FBRs.
The most prominent example of this approach being fast Fourier transforms for use with plane wave FBRs \cite{Leforestier:91a}.
In \textsc{WavePacket} we extend this approach to multidimensional problems by using a multidimensional grid formed as a direct product of one-dimensional FBR and/or DVR grids which makes our software simple and intuitive to use.
However, direct product grids lead to poor scaling as the number of degrees of freedom increases.
Hence, for large systems, better representations are needed, as implemented, for example, in MCTDH \cite{Beck2000} or in TDDVR \cite{Khan2013} software packages, which, however, require more user training for effective use.
As a consequence, \textsc{WavePacket} mainly targets small systems, with currently  up to about 4 degrees of freedom and use-cases where ease of use outweighs performance considerations.
Examples would be quantum dynamics teaching or the study of small molecules and low--dimensional model systems.

The present Part I deals with the use of \textsc{WavePacket} for closed quantum systems, i.~e., with the solution of the time-dependent and time-independent Schr\"{o}dinger equations (SE), by means of various numerical techniques.
Describing the interaction of quantum systems with external electric fields within the semi-classical dipole approximation, \textsc{WavePacket} is especially suitable for simulating experiments in photophysics or photochemistry where shaped field pulses are used to alter, and ultimately to control, the dynamics of quantum systems.
Another emphasis is on coupled SEs occuring, e.~g., when treating systems with slow and heavy degrees of freedom, such as nuclei and electrons in molecular systems.
Using \textsc{WavePacket}, the quantum dynamics of such systems can be treated beyond the Born-Oppenheimer (adiabatic) approximation, where the dynamics may be dominated by (avoided) crossings or conical intersections of potential energy curves or surfaces, respectively \cite{Baer:06a,Domcke:04a}.

The upcoming Part II will extend this discussion by describing the use of \textsc{WavePacket} for work with open quantum systems \cite{Weiss:99a,Breuer:02a}. 
In particular, it will be concerned with the Liouville-von Neumann equation, including models for dissipation and dephasing, and techniques for dimension reduction of the corresponding matrices. 
Moreover, Part II will deal with optimal control techniques which are available since the current version 5.2 of \textsc{WavePacket}.

Based on the earlier Fortran versions 2 and 3 developed during the 1990s, the \textsc{WavePacket} versions 4.x and 5.x are written in \textsc{Matlab}.
Despite the limited availability of this commercial development environment in some academic institutions, we chose that programming language because it offers several useful features.
\textsc{Matlab} is rather intuitive to use, easy to learn and offers frequently used tasks (e.g.~Fourier transforms, matrix operations) as built-in functions, allowing for fast code development.
Also, we exploit several other benefits offered by the \textsc{Matlab} environment, such as  an easy extension of core functionality through function handles.
Moreover, (animated) graphical output is readily available to the user of \textsc{WavePacket}, which is highly relevant for teaching as well as for research projects to develop a more intuitive understanding quantum dynamics.
Finally, the \textsc{Matlab} environment comes with an integrated debugger, profiler, etc., and it offers additional toolboxes for parallel computing.

Since 2008, the \textsc{Matlab} version of \textsc{WavePacket} is hosted at \textsc{SourceForge}.
This platform offers several tools that allow to manage bug tracking and feature requests, as well as the management of released versions (current version described in this work is 5.2.1).
There is also an SVN repository providing a centralized online location to control and manage distributed development of the free and open-source \textsc{WavePacket} software package; interested programmers are welcome to join its further development.
In addition, there is a large number of Wiki pages containing implementation-dependent details.
There are complete descriptions of all the \textsc{Matlab} package folders (e.~g., all implemented model potentials, initial wavefunctions, etc.), grid classes, variables, as well as input and output data files, thus serving as a reference to users.
Further information about the physical and numerical background is available on the Wiki pages of the \textsc{WavePacket} main project, along with a list of references where this software package was used already.
Also a large number of worked out demonstration examples can be found there, complete with input and output files, often including animated graphical output as MP4 files.
These examples can also be understood as a tutorial, and they are straightforward to adapt to specific needs, thus helping users to set up their own \textsc{WavePacket} simulations.

\section{qm\_setup and qm\_cleanup: Workflow of WavePacket simulations}
\label{sec::qm_setup}

A typical workflow for a \textsc{WavePacket} simulation looks as follows
\begin{verbatim}
qm_setup; 
qm_init; 
qm_propa|qm_bound|qm_movie; 
qm_cleanup
\end{verbatim}
where we assume that the \textsc{Matlab} path variable has been set appropriately. 
First, the function \texttt{qm\_setup} purges the workspace from previous calculations, closes open files, asks \textsc{Matlab} to recompile functions, and sets a multitude of default plotting parameters (see Sec. \ref{sec::qm_movie}).
In a second step, \texttt{qm\_init} is called to initialize the quantum mechanical Hamiltonian as well as the representations of wavefunctions and operators.
This function is typically provided by the user.
As will be detailed in the following Sec.~\ref{sec::qm_init}, \texttt{qm\_init} sets parameters that define the simulated system together with all necessary numeric parameters, and stores them in global variables.
It is a general policy throughout the \textsc{Matlab} version of \textsc{WavePacket} to use few, but highly structured variables to simplify book-keeping of variable names.
Moreover these variables are declared global to avoid passing values of arguments between functions as is required in some older versions of \textsc{Matlab}.
Afterwards, the functions \texttt{qm\_propa} or \texttt{qm\_bound} are invoked for a quantum propagation or for a bound-state calculation, respectively. 
These functions are the actual workhorses of the \texttt{WavePacket} package; they will be described in detail in the subsequent Secs. \ref{sec::qm_propa} and \ref{sec::qm_bound}.
Alternatively, the function \texttt{qm\_movie} can be called to produce animated graphics from stored wavefunctions generated in previous runs of \texttt{qm\_propa} or \texttt{qm\_bound}, see Sec. \ref{sec::qm_movie}.
Finally, the function \texttt{qm\_cleanup} closes the logfiles and does minor cleanup.

After the simulations have been carried out, the working directory contains several files.
The logfiles (\texttt{qm\_propa.log}, \texttt{qm\_bound.log}, or \texttt{qm\_movie.log}) hold a copy of all output that was written on the \textsc{Matlab} console. 
Depending on the parameters, there may also be various data files, image files, or movie files. 
It is also possible to save (time-dependent or time-independent) wave functions, see Sec. \ref{sec::qm_movie}.

By default, the \texttt{WavePacket} functions \texttt{qm\_propa} and \texttt{qm\_bound} not only solve the time-dependent or time-independent Schr\"{o}dinger equation, respectively, but also calculate expectation values and uncertainties for several elementary operators, such as position and momentum, potential, kinetic, and total energy.
In addition it is possible to define additional spatial projection operators, see the Wiki documentation.
Since elementary data, such as the wave functions, are easily accessible in the \textsc{Matlab} workspace, it is straightforward to calculate non-trivial physical quantities manually.

After a calculation with \texttt{qm\_propa} or \texttt{qm\_bound}, calculated data is still available until the next purge by \texttt{qm\_setup}.
In particular, the global variables holding expectation values and associated uncertainties can be imported into the current workspace with the declaration \texttt{global expect uncertain}.
Hence, the workflow explained above opens the way for \textsc{WavePacket} users to write their own scripts.
Typically one may want to run multiple calculations with similar arguments.
This can be done by modifying the \texttt{qm\_init.m} initialization function to accept one or more parameter(s), and by running \texttt{qm\_init($\cdot$)}, followed by \texttt{qm\_propa} or \texttt{qm\_bound} inside a loop, as in the following example
\begin{verbatim}
qm_setup; 
global expect
for i=1:n
    param(i) = ...
    qm_init(param(i));
    qm_propa|qm_bound;
    output(i) = expect.tot.tot(end)...
end
qm_cleanup;
clf; plot (param, output)
\end{verbatim}
After the loop, the array \texttt{param} holds a set of input parameters for each run, $1,\ldots,n$, while the array \texttt{output} holds some expectation value (here the total energy of the final wavefunction). 
The functional dependency is then available, e.~g., for a curve plot in a new figure.

\section{qm\_init: Initialize WavePacket simulations}
\label{sec::qm_init}

A closed, non-relativistic quantum mechanical system is characterized by a Hamiltonian operator $\hat{H}(R,-i\nabla_R,t)$ where $R$ is a position vector in one or more dimensions, $-i\nabla_R$ the corresponding momentum operator, and where the Hamiltonian may additionally depend on the time $t$.
Throughout the \textsc{WavePacket} software package, atomic units are used, i.~e., Planck's constant $\hbar$, the electronic mass and the elementary charge are scaled to unity. 
We emphasize that \textsc{WavePacket} can be employed not only for a single ($\nu=1$) but also for several ($\nu>1$) coupled Schr\"{o}dinger equations in which case the Hamiltonian becomes a $\nu\times\nu$ operator matrix. 
The latter case arises naturally within the Born-Huang description of molecular quantum dynamics, where $R$ are the nuclear degrees of freedom and where $\nu$ is the number of electronic states involved in a close coupling calculation.

\subsection{Hamiltonian operator}
\label{sec::qm_init::ham}
The structure of the Hamiltonian for any \textsc{WavePacket} simulation must generally have the form
\begin{equation}
\hat{H}(R,-i\nabla_R,t)=\hat{T}(R,-i\nabla_R)+\hat{V}(R)-i\hat{W}(R)-F(t)\cdot\hat{\mu}(R)
\label{eq:Hamilt}
\end{equation}
with the following symbols:
\begin{itemize}
\item $\hat T(R,-i\nabla_R)$ is the kinetic energy operator whose actual form is often tied to the underlying finite basis representation, see Sec. \ref{sec::qm_init::dvr}.
For $\nu>1$, the corresponding operator matrix must be diagonal and have identical operators for all diagonal elements, i.~e., \textsc{WavePacket} always requires a so-called diabatic formulation of the Hamiltonian, see also Sec. \ref{sec::qm_propa::cc}. 

\item $\hat V(R)$ is a potential energy function governing the dynamics. For applications in molecular sciences, it has to be generated by quantum-chemical calculations or has to be based on empirical potential functions. The potential becomes a symmetric matrix of real-valued functions for $\nu>1$ within the diabatic representation. 
Note that the diabatization of arbitrary quantum-chemical data is not trivial and outside the scope of \textsc{WavePacket}.

\item $-i\hat W(R)$ is a negative imaginary potential (NIP), optionally used in \textsc{WavePacket} to smoothly absorb wavefunctions near the boundaries. 
Similar to the kinetic energy, this absorber becomes a diagonal matrix with identical elements for $\nu>1$.

\item $F(t)$ is an external electric field that can be optionally used to control the dynamics of the quantum system under consideration \cite{Zewail2000,Grossmann2008,Sundstrom2008}. 
The current implementation of \textsc{WavePacket} is based on the semi-classical treatment within the electric dipole approximation where the electric field couples to the system through its dipole operator $\hat{\mu}(R)$, see Sec. \ref{sec::qm_propa::vib}.

\item $\hat\mu(R)$ is a coordinate dependent dipole moment function of the quantum system under consideration. 
The dipole moment becomes a real symmetric matrix for $\nu>1$ , again within the diabatic representation used throughout \textsc{WavePacket}. 
In that case, the diagonal and off-diagonal elements are referred to as permanent and transition dipole moments $\mu_p(R)$ and $\mu_t(R)$, respectively. 
\end{itemize}


The \textsc{WavePacket} program package comes with a rather large choice of model functions for the coordinate-dependence of the potential function $V(R)$, dipole moment functions $\mu(R)$, as well as the negative imaginary potential $W(R)$.
These `libraries' are realized as \textsc{MATLAB} package folders named here \texttt{+pot, +dip, +nip}, respectively.
For further details, see the Reference Manual on the Wiki pages.
Additional model functions can be easily written to simulate also photophysical/photochemical situations not covered by the built-in functions.
As an alternative, $V(R)$ and $\mu(R)$ can be specified in terms of a Taylor series, or they can be given as tabulated values through formatted data files, which are then interpolated inside \textsc{WavePacket}.

\subsection{Example: Hamiltonian of a Morse oscillator}
\label{sec::qm_init::morse}

As an example, let us consider a one-dimensional Morse oscillator with potential energy 
\begin{equation}
	V(R) = D_e \left(1-\mathrm{e}^{-\alpha(R-R_e)}\right)^2
	\label{eq::morse}
\end{equation}
with dissociation energy $D_e$, equilibrium position $R_e$ and range parameter $\alpha>0$.
We let the oscillator interact with an external electric field through a dipole moment modeled by a Mecke function \cite{Mecke:50a}
\begin{equation}
	\mu(R)=  q_0 R \mathrm{e}^{-R/R_0}
		\label{eq::mecke}
\end{equation}
where $q_0$ and $R_0$ are charge and distance parameters. 
Moreover, in time-dependent simulations we use a (truncated) power function as negative imaginary potential to smoothly absorb the wavefunction outside the interval [$R_\mathrm{min},R_\mathrm{max}$] 
\begin{eqnarray}
	W(R) &\propto& \Theta(R_\mathrm{min}-R)(R_\mathrm{min}-R)^\beta \nonumber \\
	&+& \Theta(R-R_\mathrm{max})(R-R_\mathrm{max})^\beta
\end{eqnarray}
with $\beta>0$ and where $\Theta$ is the Heavyside function.
Although this system is rather simple, it is sufficient to serve as a standard example throughout most of this work. 
Generalizations to more dimensions and/or more complex scenarios can be found in the demo examples available in the Wiki documentation of \textsc{WavePacket} at SourceForge.

All settings concerning the Hamiltonian, as well as further settings required for other parts of WavePacket explained below have to be made by the user. 
Most conveniently this can be achieved with a user-defined function, which we suggest to call \texttt{qm\_init}. 
For the above example, the beginning of this function could read as follows 
{\small\begin{verbatim}
global hamilt;

hamilt.pot.handle      = @pot.morse;
hamilt.pot.params.d_e  = 0.1994; 
hamilt.pot.params.r_e  = 1.821; 
hamilt.pot.params.alf  = 1.189;

hamilt.dip.handle      = @dip.mecke;
hamilt.dip.params.r_0  = 1.1338359;
hamilt.dip.params.q_0  = 1.6343157;

hamilt.nip.handle      = @nip.power; 
hamilt.nip.params.exp  = 4; 
hamilt.nip.params.min  = 1.0;
hamilt.nip.params.max  = 6.0;
\end{verbatim}}

The first line makes the structured global variable \texttt{hamilt} available inside the function.
The choices of (Morse) potential function, (Mecke) dipole function, and (power) absorber function are realized with function handles, i.e., references to functions indicated by the \texttt{@}-sign, with the corresponding functions taken from the package folders \texttt{+pot, +dip, +nip}, respectively.
The use of function handles allows easy customization; in particular, they can also reference any user-supplied function.
The values of the parameters are chosen such as to resemble an OH oscillator inside a water molecule in its electronic ground state \cite{BSchmidt:14}.

\subsection{Finite basis and discrete variable representations}
\label{sec::qm_init::dvr}
All \textsc{WavePacket} codes are based on the standard approach in numerical quantum mechanics employing expansions of  wave function(s), $\Psi(R)$, and the Hamiltonian $H$ into a truncated orthonormal basis $P_n$.
As an example, consider a one-dimensional wave function with $R \in [-1,1]$ that is expanded into normalized Legendre polynomials $P_n$.
Neglecting contributions from dipole terms and negative imaginary potentials for ease of notation, the expansion yields
\begin{eqnarray}
	\Psi(R) &=& \sum_{n=0}^{N-1} \psi_n P_n(R) \nonumber \\
	\hat H = \hat T + \hat V &=& \sum_{m,n=1}^{N-1} (T_{mn}+V_{mn}) |P_m \rangle \langle P_n|
	\label{eq::dvr::fbr}
\end{eqnarray}
which is called a finite basis representation (FBR) \cite{Light:85a,Light:00a}.
Often, the basis can be chosen such that the kinetic energy matrix $T_{mn}$ is diagonal or at least very sparse, while in general the potential matrix $V_{mn}$ is not.
To avoid the cumbersome and expensive construction of $V_{mn}$, we can search for a representation that diagonalizes the potential matrix.
To this end, we can use the Gauss-Legendre quadrature, which states that there are $N$ weights, $w_i$, and points, $R_i$, such that
\begin{equation}
\label{eq::dvr::quadrature}
	\int_{-1}^1 f(R) dR = \sum_{i=1}^N w_i f(R_i)
\end{equation}
is exact for all polynomials $f(R)$ up to degree $2N-1$.
For our purposes, we can substitute $f$ by $\Psi$, which is a polynomial by definition (\ref{eq::dvr::fbr}).
The $N$ values of the wave function at the grid points $\Psi(R_i)$ then form the discrete variable representation (DVR) \cite{Light:85a,Light:00a}.
They contain the same information as the FBR coefficients $\psi_n$, since the transformations
\begin{eqnarray}
\Psi(R_i) = \sum_{n=0}^{N-1} \psi_n P_n(R_i)
\end{eqnarray}
with coefficients
\begin{eqnarray}
\psi_n = \langle{P_n}|{\Psi}\rangle = \int_{-1}^1 P_n^\ast(R) \Psi(R) dR = \sum_{i=1}^N w_i P_n^\ast(R_i) \Psi(R_i)
\end{eqnarray}
are exact within the truncated basis (\ref{eq::dvr::fbr}).

We can now apply the potential energy operator onto the wave function, $\Psi'(R) = \hat V \Psi(R)$, and determine the new FBR coefficients
\begin{equation}
\label{eq::dvr::boring}
\psi'_m = \langle{P_m}|{\Psi'}\rangle = \langle P_m| \hat V |\Psi\rangle = \int_{-1}^1 P_m^\ast(R) V(R) \Psi(R) dR	.
\end{equation}
Note that the FBR (\ref{eq::dvr::fbr}) is only a good representation if also $\Psi'(R) = V(R)\Psi(R)$ is a polynomial of degree $<N$.
However, in this case, the integrand is a polynomial of degree $\leq 2N-1$, and we can use the quadrature (\ref{eq::dvr::quadrature}) to get
\begin{equation}
	\psi'_m = \sum_{i=1}^N w_i P_m^\ast(R_i) V(R_i) \Psi(R_i),
	\end{equation}
or with some more algebra 
\begin{equation}
\label{eq::dvr::dvr_method}
	\Psi'(R_i) = V(R_i) \Psi(R_i).
\end{equation}
Hence, in the DVR the potential energy matrix is diagonal, that is, acts through elementwise multiplication with a vector composed of the potential at the grid points.

In summary, the DVR basis is often easier to use than the FBR basis.
In particular, operators of the type $\hat A = A(R)$ are easily applied through Eq.~(\ref{eq::dvr::dvr_method}) without an explicit matrix representation.
Also, the DVR coefficients of an initial wave function $\Psi_0(R)$ are trivially obtained as $\Psi_0(R_i)$, while the FBR would require the evaluation of scalar products $\langle P_n|\Psi_0\rangle$.

Currently, the \textsc{WavePacket} program package implements three types of DVR schemes:
\begin{enumerate}
	\item 
Schemes based on Legendre and Gegenbauer polynomials are extending the Gauss quadrature (\ref{eq::dvr::quadrature}) with weight functions, to be included in the FBR basis.
Legendre polynomials $P_l$ and the more general Gegenbauer polynomials $C_n^{(\alpha)}$, can be used to express spherical harmonics as
\begin{eqnarray}
	Y_{l0}(\theta, \phi) &\propto& P_l(\cos \theta) \nonumber \\
		Y_{lm}(\theta, \phi) &\propto& (\sin \theta)^{\frac{m}{2}} C_{l+m}^{(\frac{1}{2}-m)}(\cos \theta) \mathrm{e}^{im\phi}
	.
\end{eqnarray}
Hence, the corresponding Gauss-Legendre DVR finds natural use in rotational problems where the potential $V=V(\theta)$  is of azimuthal symmetry, i.~e., where the quantum number $m$ is conserved.
In such rotational problems, the relevant kinetic operator is usually $\frac{1}{2MR^2}\hat{L}^2$ expressed in terms of the angular momentum operator $\hat L$ and mass $M$, and is turned on by default when using such a Gauss-Legendre DVR, where $R$ can either be a constant or be supplied from another radial degree of freedom.

\item
Up to a weight function, Hermite polynomials describe harmonic oscillator eigenstates.
Hence, they are suitable for Cartesian coordinates, and the corresponding Gauss-Hermite DVR becomes more efficient the closer the potential $V(R)$ is to its harmonic approximation.
Hence, this scheme is most suitable for a vibrational problem.
The kinetic energy operator that is used by default is $-\frac{1}{2M}\frac{\partial^2}{\partial R^2}$ with mass $M$.

\item
\textsc{WavePacket} also contains the Fourier-method \cite{Kosloff:83a}, where the FBR corresponds to an expansion into plane waves with implicit periodic boundary conditions, and where the DVR is obtained by a discrete Fourier transform.
Even though plane waves often provide a poor representation of the wave function and require many basis functions for convergence they are often used in quantum dynamics simulations.
This is because the transformations between FBR and DVR can be realized by means of the fast Fourier transformation (FFT), the effort of which scales much better than a matrix multiplication.
Hence, the FFT method is especially suitable, e.~g., for simulations of molecular dissociation \cite{BSchmidt:14,BSchmidt:59}, association \cite{BSchmidt:19,BSchmidt:20,BSchmidt:26,BSchmidt:37}, or ionization \cite{BSchmidt:23,BSchmidt:44} in the wavepacket picture.
Choosing a Fourier DVR for some degree of freedom by default also uses the same Cartesian kinetic operator as for the Gauss-Hermite DVR.
\end{enumerate}

It should also be noted that there is a large body of literature on quadrature schemes for a variety of other, more specialized basis functions \cite{Light:85a,Manolopoulos:88a,Muckerman:90a,Light:00a}.

\subsection{Example: FFT grid representations}
\label{sec::qm_init::fft}

Here we return to the example of the Morse oscillator introduced above and add 
a few lines to \textsc{Matlab} initialization function \texttt{qm\_init} that define the grid DVR and the associated kinetic operator
{\small\begin{verbatim}
global space;
space.dof{1}       = grid_fft; 
space.dof{1}.mass  = 1728.539;
space.dof{1}.n_pts = 256;
space.dof{1}.x_min =  0.7;
space.dof{1}.x_max = 10.0;
\end{verbatim}}
Again, the first line serves to make the structured variable \texttt{space} globally accessible within \textsc{WavePacket}.
The second line indicates our choice of using an FFT--based plane wave FBR. The further lines specify the necessary parameters, here the number of grid points as well as the lower and upper bounds of the grid.
Also, the default kinetic energy operator, here the standard Cartesian kinetic energy $-\frac{1}{2M}\frac{\partial^2}{\partial R^2}$, is set up, needing the (reduced) mass $M$ to be specified.
While traditional FFT routines require the number of grid points to be an integer power of two, the "Fastest Fourier Transform in the West" (FFTW) routines used inside \textsc{MATLAB} are more flexible; however, be aware that FFTW works best if the number of grid points can be decomposed into as small as possible prime factors. Hence, integer powers of two are ideal but by no means mandatory.

We note in passing that the kinetic energy operators required for a quantum mechanical system can differ from the defaults provided with each grid.
To support such cases, it is possible to turn off the default kinetic energy operators, and to supply custom implementations, although these are slightly more involved than for example custom potential functions.
Two variants occuring in triatomic systems, namely the kinetic energy operator for internal Jacobi coordinates and for triatomic systems with fixed bending angle are supplied as examples.

\subsection{Multi-dimensional grid representations}
For a multi-dimensional simulation, the cell vector \texttt{space.dof}\{$\cdot$\} has additional entries that describe the FBR bases and the DVR grids along each degrees of freedom.
Technically, each of the entries \texttt{space.dof} is an instance of a grid class (\texttt{grid\_legendre, grid\_hermite, grid\_fft}) that encapsulates one of the DVR schemes discussed in Sec.~\ref{sec::qm_init::dvr}.
For historical reasons, these use the old (prior to version 7.6) \textsc{MATLAB} implementation for object-oriented programming.
The grid classes contain the functionality for the setup of the grid points and weights, for transformations between the FBR and the DVR, and for the evaluation of the corresponding default kinetic operator.

The one-dimensional grids are stored as the entries of a cell vector \texttt{dvr.grid\_1D} (i.e., \texttt{dvr.grid\_1D\{$k$\}}($\cdot$) for the $k$-th degree of freedom).
Corresponding multidimensional grids, \texttt{dvr.grid\_ND}\{$\cdot$\}$(\cdot,\cdot,\ldots)$, are then constructed from these one-dimensional grids using the \textsc{Matlab}'s \texttt{ndgrid} function.
In close analogy, \textsc{WavePacket} uses tensors, i.~e., multi-dimensional arrays, to store all other grid-representable data, such as wavefunctions, potentials, or any other operators diagonal in the DVR or FBR.
For example, the coordinate representation of the $m$-th wavefunction in a close-coupling calculation is stored in the tensor \texttt{psi.dvr.grid\_ND}\{$m$\}$(\cdot,\cdot,\ldots)$, which is of size $(N_1, N_2, \ldots)$ with $N_k$ being the numbers of grid points along each degree of freedom $k$.

\section{qm\_propa: Solving the time-dependent Schr\"{o}dinger equation}
\label{sec::qm_propa}
Formulated as a partial differential equation, the time-dependent Schr\"{o}dinger equation (TDSE) describes the evolution of a quantum system in terms of a complex-valued wave functions $\Psi(R,t)$,
\begin{equation}
i\frac{d}{dt}\Psi(R,t) = \hat H(R, -i\nabla_R, t) \Psi(R, t), \quad \Psi(R,t=0)=\Psi_0(R).
\label{eq::tdse::eom}
\end{equation}
Here $R$ is a one- or multi-dimensional position vector, $t$ is the time, and the Hamiltonian $\hat H$ from Eq. (\ref{eq:Hamilt}) can be in general time-dependent; it becomes an operator matrix for close coupling calculations, $\nu>1$ (see our remarks in Sec.~\ref{sec::qm_init} on the use of diabatic representations in \textsc{WavePacket}).
From the solution $\Psi(R,t)$, expectation values of observables $\hat O$ can be calculated as 
\begin{equation}
\langle \hat O \rangle(t) = \langle\Psi(t)|\hat O|\Psi(t)\rangle .
\label{eq::tdse::expect}
\end{equation}
where the operator $\hat O$ may be matrix valued for $\nu>1$.
In this case, \textsc{WavePacket} calculates expectation values both for each component of the operator $\hat O$ as well as for the complete quadratic form.
In \textsc{WavePacket}, the TDSE (\ref{eq::tdse::eom}) is solved using the function \texttt{qm\_propa} which comprises three classes of solvers: 
Runge-Kutta methods, the split operator scheme, and polynomial propagators.
In all cases, the principal idea is to approximate the short-time propagation
\begin{equation}
	\Psi(t+\Delta t) = \hat U_{t+\Delta t, t} \Psi(t)
\end{equation}
where the time evolution operator is defined as
\begin{equation}
\label{eq::tdse::propagator}
	\hat U_{t+\Delta t,t} = \mathcal{T} \mathrm{e}^{-i \int_t^{t+\Delta t} \hat H(\tau) d\tau} = \mathrm{e}^{-i \hat H \Delta t}
\end{equation}
with the formal time ordering operator $\mathcal{T}$, and where the second equality holds only for time-independent Hamiltonians, but can be considered an approximation for sufficiently short time steps $\Delta t$ otherwise.

\subsection{Second-order differencing}
\label{sec::qm_propa::sod}
One of the simplest Runge-Kutta methods is the second-order differencing method \cite{Askar:78a}, a time-symmetrized, symplectic Euler integration
\begin{equation}
\label{eq::tdse::sod}
		\Psi(t+\Delta t) = \Psi(t-\Delta t) -2 i \Delta t \hat H(t) \Psi(t)  + \mathcal{O}(\Delta t^3)
	.
\end{equation}
which is similar in spirit to the Verlet-St\"{o}rmer scheme in classical mechanics.
The differencing method approximately preserves the norm of the wave function up to $\mathcal{O}(\Delta t^4)$, is easy to implement, and also works for time-dependent Hamiltonians.
However, the errors necessitate small time steps thus making this scheme inferior to other methods \cite{Leforestier:91a}.

\subsection{Split operator}
\label{sec::qm_propa::spo}
The split operator scheme \cite{Fleck:76a,Feit:82a} works by independently exponentiating the summands of the Hamiltonian (\ref{eq:Hamilt}).
In lowest orders, this leads to the first-order Trotter splitting,
\begin{equation}
\label{eq::tdse::trotter}
	\hat U_{t+\Delta t,t} =
	\mathrm{e}^{-i \hat T \Delta t} 
	\mathrm{e}^{+i F(t) \cdot \hat\mu_t \Delta t} 
	\mathrm{e}^{+i F(t) \cdot \hat\mu_p \Delta t} 
	\mathrm{e}^{-i \hat V \Delta t} 
	+ \mathcal{O}(\Delta t^2),
\end{equation}
and the second-order Strang splitting,
\begin{eqnarray}
\label{eq::tdse::strang}
	\hat U_{t+\Delta t,t} &= &
		\mathrm{e}^{-i \hat V \Delta t/2} 
		\mathrm{e}^{+i F(t+\Delta t) \cdot \hat\mu_p \Delta t/2} 
		\mathrm{e}^{+i F(t+\Delta t) \cdot \hat\mu_t \Delta t/2}
		\mathrm{e}^{-i \hat T \Delta t}
		\nonumber \\
	&\times & 
	\mathrm{e}^{+i F(t) \cdot \hat\mu_t \Delta t/2} 
	\mathrm{e}^{+i F(t) \cdot \hat\mu_p \Delta t/2} 
	\mathrm{e}^{-i \hat V \Delta t/2} 
		+ \mathcal{O}(\Delta t^3).
\end{eqnarray}
The schemes exploit the fact that the kinetic energy $\hat T$ usually has a diagonal matrix representation in the chosen FBR, while the potential $\hat V$, as well as (permanent and transition) dipole moments $\hat \mu_p, \hat \mu_t$ are diagonal in the DVR.
For diagonal operators, the exponentiated operator is directly obtained from an elementwise exponentiation of the matrix elements.
Hence, in practice, the speed of execution is determined by the efficiency of the transformations between the representations needed for calculating the single exponentials.
Also, we point out that the terms for the dipole moments need to be exponentiated on every time step; for the transition dipole moment, this may require a transformation into another representation where the dipole moment is diagonal in the close coupling scheme.
Also note that the negative imaginary potential is only applied after larger time steps which is chosen for consistency with the Chebychev method described in Sec. \ref{sec::qm_propa::cheb}.

Both split-operator schemes lead to unitary time evolution, that is, they guarantee that the norm of the wave function is conserved.
Though extensions to higher orders are possible \cite{Bandrauk:92a}, they are not included in \textsc{WavePacket}
because the number of exponential terms, and hence the computational effort, grows quickly with increasing order.

\subsection{Chebychev method}
\label{sec::qm_propa::cheb}
The idea of polynomial propagators is to expand the time evolution operator in a truncated series of polynomials $P_n$.
They can be fast and highly accurate \cite{Leforestier:91a}, but typically require time-independent Hamiltonians, and rather large time steps for efficiency.
The general approach can be written as
\begin{equation}
	\hat U_{t+\Delta t,t} = \mathrm{e}^{-i \hat H \Delta t} \approx \sum_{n=0}^N c_n P_n(-i \hat H \Delta t)	.
\end{equation}
One scheme for the determination of optimal coefficients $c_n$ can be obtained by rewriting this equation in the eigenstate basis of $H$,
\begin{equation}
	\hat U_{t+\Delta t,t} = \sum_k \mathrm{e}^{-i E_k \Delta t} |\varphi_k\rangle\langle\varphi_k| 
		\approx \sum_k \Bigl( \sum_{n=0}^N c_n P_n(-i E_k \Delta t) \Bigr) |\varphi_k\rangle\langle\varphi_k|
	.
\end{equation}
The polynomial series approximates the terms $\exp(-i E_k \Delta t)$, which is formally similar to an expansion of an exponential in a polynomial basis.
Hence, with $x = -i E_k \Delta t$, we can determine the coefficients as
\begin{equation}
\label{eq::tdse::coefficients}
	c_n = \frac{\langle P_n | \mathrm{e}^{x}\rangle}{\|P_n\| \|\mathrm{e}^{x}\|}
	= \frac{1}{\|P_n\| \|\mathrm{e}^{x}\|} \int_{x_\mathrm{min}}^{x_\mathrm{max}} P_n^\ast(x) \mathrm{e}^{x} dx	.
\end{equation}
One of the most widely-used polynomial propagators is based on the use of Chebychev polynomials \cite{Tal-Ezer:84a,Chen:99b} where a rescaled Hamiltonian with eigenvalues in the range $[-1,1]$ is used.
Also the rescaled time step 
\begin{equation}
\alpha = \frac{\Delta E \Delta t}{2}
\label{eq::tdse::alpha}
\end{equation} 
is kept as an explicit parameter in the coefficients, $c_n =c_n(\alpha)$, where $\Delta E=E_\mathrm{max}-E_\mathrm{min}$ is the spectral range of the DVR of the Hamiltonian.
The coefficients turn out to be proportional to Bessel functions of the first kind, $c_n(\alpha) \propto J_n(\alpha)$, which decay exponentially for orders $n>\alpha$, so that almost arbitrary accuracy can be reached with only moderate effort.

However, Chebychev propagators only work for Hermitian Hamiltonians, because the integration in (\ref{eq::tdse::coefficients}) extends only over purely imaginary $x \propto -i E_k$.
While it is possible to construct a similar propagator that works for propagation in imaginary time \cite{Kosloff:86a}, see Sec. \ref{sec::qm_bound::relax}, Chebychev propagators cannot be applied to systems with complex eigenvalues, which appear, e.~g., for open quantum systems or for absorbing boundary conditions.
Some of these limitations, however, can be overcome with other polynomial propagators, see, e.~g., Ref.~\cite{Huisinga:99a}.
Also, we note that we can still apply the absorbing boundary conditions explicitly outside of the propagator after each time step $\Delta t$.

\subsection{Example: Fractional revivals of a Morse oscillator}
\label{sec::qm_propa::rev}
As an example on the use of the \texttt{qm\_propa} function , we come back to the  Morse oscillator introduced above.
The specification of an initial wavefunction $\Psi(R,t=0)=\Psi_0(R)$ requires the following settings inside global structure \texttt{psi}, preferentially within the initialization file \texttt{qm\_init}:
\begin{verbatim}
global psi
psi.init.dof{1}.handle = @wav.morse;
psi.init.dof{1}.d_e    = 0.1994;
psi.init.dof{1}.r_e    = 1.44;
psi.init.dof{1}.alf    = 1.189;
\end{verbatim}
The choice of the initial state, here an eigenfunction of the Morse potential, proceeds via a function handle that points to the function \texttt{morse} inside the \texttt{+wav} package folder.
Note that we choose the ground state wavefunction of the underlying Morse potential itself, but with its equilibrium shifted (here from 1.821 to 1.44), similar to a shifted (coherent) state of a harmonic oscillator \cite{Glauber:63a,Schleich:01a}.

In case of multi-dimensional simulations, initial wavefunctions can be constructed as direct products of one-dimensional wavefunctions specified by the entries of \texttt{psi.init.dof\{$\cdot$\}}.
\textsc{WavePacket} comes with several standard functions such as Gaussian wavepackets, eigenstates of harmonic and Morse oscillators, or pendular states that are detailed on the Wiki.
Alternatively, also fully correlated Gaussian packets (or linear combinations thereof), or tabulated values are available.

The details of the numeric propagator and the time discretization are set in the global structure \texttt{time}
\begin{verbatim}
global time
time.propa.handle = @ket.cheby_real;
time.propa.params.precision = 10^-8;
time.main.stop  = 100;
time.main.delta  = 76.8237;
\end{verbatim}
where the Chebychev propagator in real time from package folder \texttt{+ket} is chosen and the polynomial expansion is truncated automatically once the coefficients fall below $10^{-8}$.
Here the simulation proceeds in 100 main steps of size 76.82 each ($\alpha=76.82$ for $\Delta E$ truncated to 2, see Eq. (\ref{eq::tdse::alpha})) which requires the evaluation of 104 polynomials. 

\begin{figure}
	\caption{Simulation of wavepacket dynamics in a Morse oscillator using \texttt{qm\_propa}.
	The snapshots show Wigner representations along with position and momentum densities as marginals, with color coding corresponding to complex phase of wavefunctions. Upper panel: After 5 time steps ($t=384.12$) the dephasing begins. 
	Lower panel: After 100 time steps ($t=7682.37$) there is a fractional revival of order 1/2.}
	\includegraphics[width=10cm]{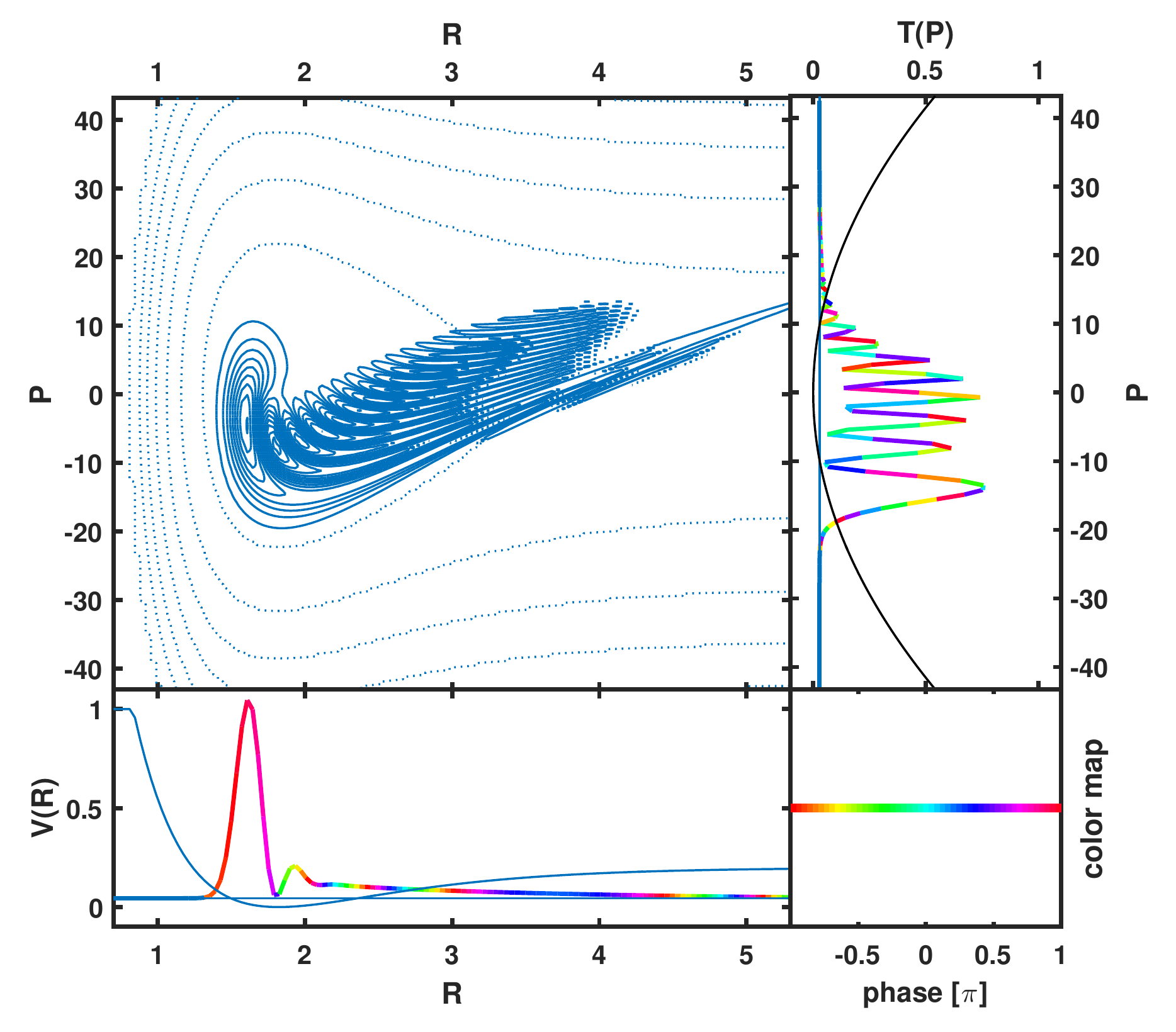}\\[1cm]
	\includegraphics[width=10cm]{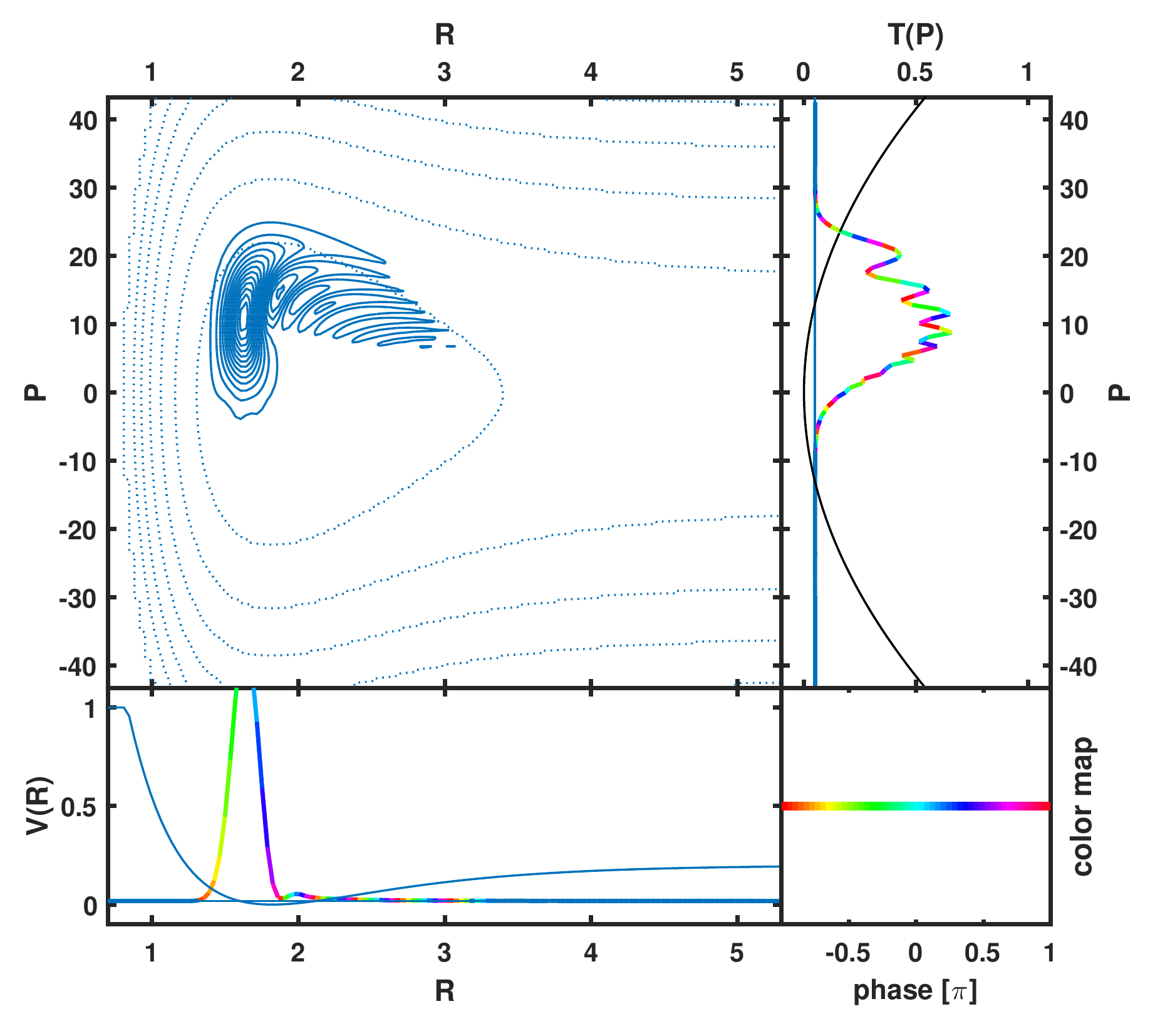}
	\label{fig:revival}
\end{figure}

Soon after the start of this \textsc{WavePacket} propagation, position and momentum wavefunctions become essentially unstructured, with the Wigner distribution smeared over the energetically accessible portion of phase space, as shown in the upper panel of Fig.~\ref{fig:revival}.
Also some loss of the probability density is observed which is due to the absorbing boundaries.
At much longer times, however, regular, periodic motion is resumed.
 This rephasing can be explained in terms of full and fractional revivals, see e.~g. Refs \cite{Leichtle:96a,Leichtle:96b}.
The lower panel of Fig.~\ref{fig:revival} shows an almost complete restoration of the initial shape of the wavepacket which occurs at one half of the revival time, here for $t=7682$.

\subsection{Vibrational excitation of a diatomic molecule}
\label{sec::qm_propa::vib}
In another example, we shall discuss quantum dynamics driven by external fields.
In agreement with typical experimental scenarios, WavePacket assumes the time-dependence of the electric field to be given as a sum of pulses centered around $t=\tau$, with amplitude $F_0$ and phase $\phi_0$
\begin{equation}
	F(t)=F_0 g(t-\tau) \cos\left[\omega(t-\tau)\cdot(t-\tau)+\phi_0\right]
	\label{eq::tdse::pulse}
\end{equation}
The fields oscillate with carrier frequency 
\begin{equation}
 \omega(t)=\omega_0+\gamma t+\frac{1}{2}\delta t^2
\end{equation}
with constant frequency $\omega_0$ or optionally with linear ($\gamma$) and quadratic ($\delta$) time dependence ("`chirp"').
There is a choice of different model functions for the pulse shape or envelope, $g(t)$, for which we refer the reader to the Wiki documentation.
As an alternative to the use of model functions, there is also the possibility of reading tabulated values of $F(t)$ from an input data file which are then interpolated.
This is especially important when dealing with pulses obtained from optimal control theory as we will discuss in Part II.

\begin{figure}
	\caption{Simulation of vibrational excitation of a Morse oscillator using \texttt{qm\_propa}. 
	Upper panel: Electric field of a 1 ps infrared laser pulse, see Ref.~\cite{BSchmidt:14}. 
	Lower panel: Induced population dynamics shows vibrational ladder climbing in a 5-photon excitation with very high vibrational selectivity.}
	\includegraphics[width=10cm]{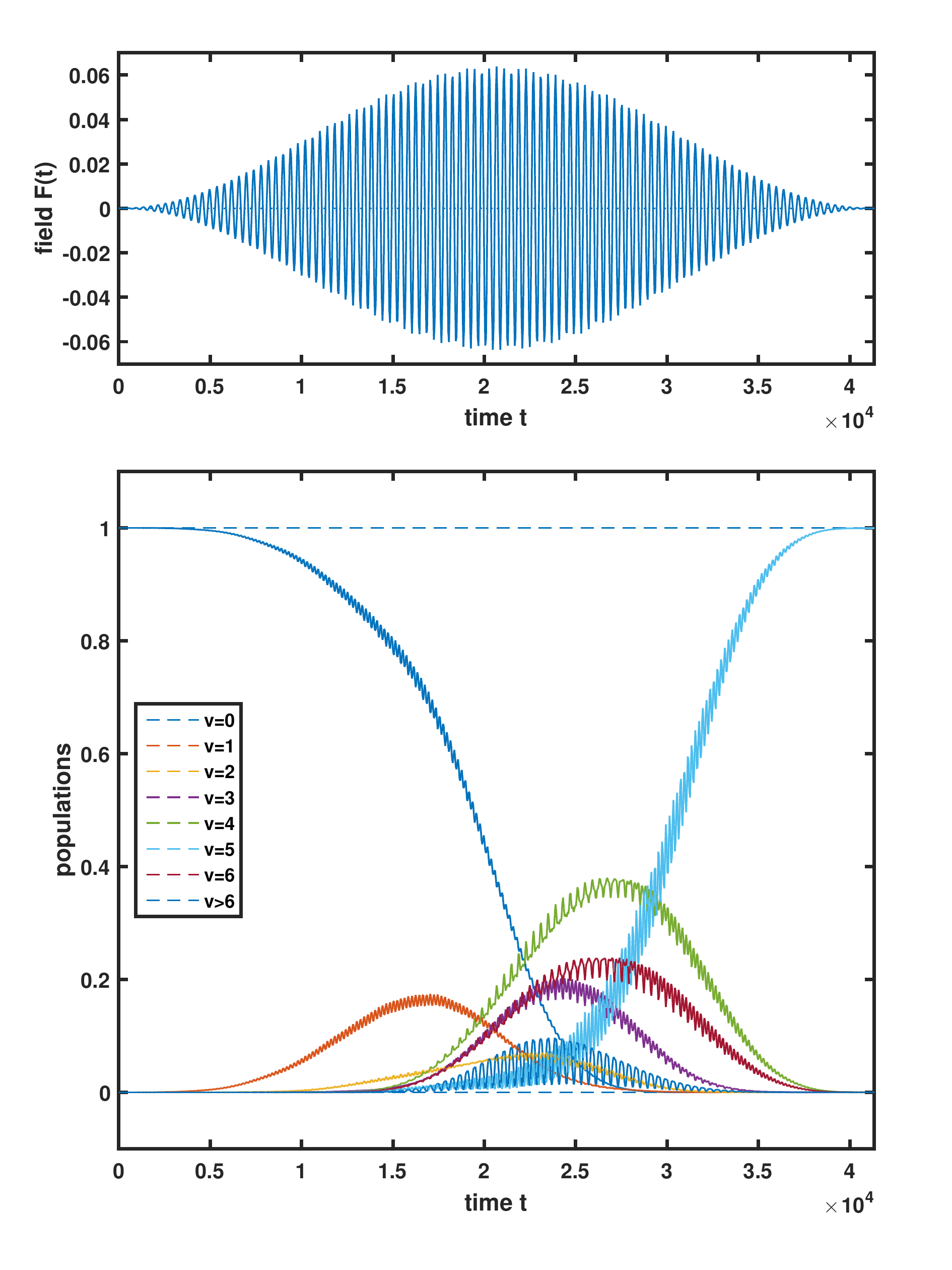}
	\label{fig:excite}
\end{figure}

As specific example, let us consider the vibrational excitation of a diatomic molecule, which is modeled by the Morse oscillator potential and Mecke dipole function introduced in Secs. \ref{sec::qm_init::morse}, \ref{sec::qm_init::fft}.
We assume the system to be initially in its ground state, similarly as in Sec.~\ref{sec::qm_propa::rev} but with its equilibrium unshifted, i.e.~\texttt{psi.init.dof\{1\}.r\_e = 1.821}.    
The time-dependence of the electric field can be specified by the following lines in \texttt{qm\_init.m}
\begin{verbatim}
time.efield.shape = 'sin^2';
time.efield.delay = 500 * 41.341373;
time.efield.fwhm  = 500 * 41.341373;
time.efield.ampli = 328.5 / 5142.2064;
time.efield.frequ = 3424.19/219474.63;
\end{verbatim}
where we have chosen a single sin$^2$ shaped laser pulse \cite{Barth2009} with 500 fs duration (FWHM), a maximum field strength of $F_0=328.5$ MV/cm and a constant carrier frequency $\omega_0=3424.19$ cm$^{-1}$, see upper  panel of Fig.~\ref{fig:excite}.
For the numeric propagation, we choose a split operator scheme with error $\propto (\Delta t)^3$, i.~e., the Strang scheme (\ref{eq::tdse::strang})
\begin{verbatim}
time.propa.handle = @ket.splitting; 
time.propa.params.order = 3
time.main.delta = 413.41; 
time.main.stop  = 100; 
time.sub.n      = 100; 
\end{verbatim}
Note that \textsc{WavePacket} uses two time steps. After each of the main time steps, here, 10 fs $\approx$ 413.41 a.u., all expectation values are calculated and printed to the \textsc{MATLAB} console and the logfile. 
Optionally, also the absorbing boundary conditions are applied, a frame will be added to the animated graphics file, and the wave function may be saved. 
The definition of the main time step is sufficient for Chebychev propagators, which require rather long propagation steps for efficiency.
For other propagators, such as the second order differencing scheme of Eq. (\ref{eq::tdse::sod}) or the splitting schemes of Eqs. (\ref{eq::tdse::trotter}) and (\ref{eq::tdse::strang}), the main time steps can be divided into shorter ``substeps'', which are then actually used as propagation step $\Delta t$ for the evaluation of the short time propagators
The oscillatory field used in this example requires about 100 substeps per main time steps, yielding a propagation step of $10$ fs $/100=0.1$ fs.
As can be seen in the lower panel of Fig.~\ref{fig:excite}, after 100 main steps or at $t=41341$, more than 99.9\% of the population has been transferred from the $v=0$ to the $v=5$ level of the Morse oscillator.
Hence, the infrared laser pulse has achieved a vibrationally state selective 5-photon excitation \cite{BSchmidt:14}.

Note that in \textsc{WavePacket} the field $F(t)$ as well as the dipole operator $\hat{\mu}(t)$ can have one or two ($x$ and $y$) Cartesian components thus allowing to model polarization effects.
An example can be found in Ref. \cite{BSchmidt:26} where such effects are used to control the branching of photoassociation in the $X ^2\Pi$ versus $A ^2\Sigma^+$ states of the OH radical studied as an example above.

If the shape function $g(t)$ in Eq. (\ref{eq::tdse::pulse}) varies slowly in time, it can be useful to transform to the Floquet picture of light-dressed states \cite{Guerin:97a,BSchmidt:34,Guerin:03a} which is also implemented.
Apart from numerical advantages \cite{BSchmidt:41}, this picture naturally offers interpretations of quantum dynamics of externally driven systems within an adiabatic approximation \cite{Breuer:89a,Drese:99a,BSchmidt:59}.

\subsection{Close coupling schemes}
\label{sec::qm_propa::cc}
When dealing with multiple ($\nu>1$) coupled Schr{\"o}dinger equations, the variable \texttt{hamilt.coupling.n\_eqs} has to be set to their number $\nu$. 
Then wavefunctions are stored in a cell vector, \texttt{psi.dvr.grid\_ND}\{$\cdot\}(\cdot,\cdot,\ldots)$ and diabatic potentials are stored in a cell matrix, \texttt{hamilt.pot.grid\_ND}\{$\cdot,\cdot\}(\cdot,\cdot,\ldots)$.
In both cases each of the cell entries, in turn, is a multi-dimensional array, stemming from the DVR of the position coordinates.
Note that cell arrays offer additional flexibility due to the fact that individual entries of cell arrays may also be empty. 
This is conveniently used, e.~g., if certain couplings are forbidden due to symmetry restrictions.

\begin{figure}
	\caption{Non-adiabatic quantum dynamics simulation using \texttt{qm\_propa}. For a single Gaussian starting in the upper left, the snapshots are taken shortly after having passed the transition region. 
	Upper panel: Two linear potential energy surfaces intersecting each other along a linear seam in a diabatic representation. 
	Lower panel: Conical intersection in an adiabatic representation.}
		\includegraphics[width=10cm]{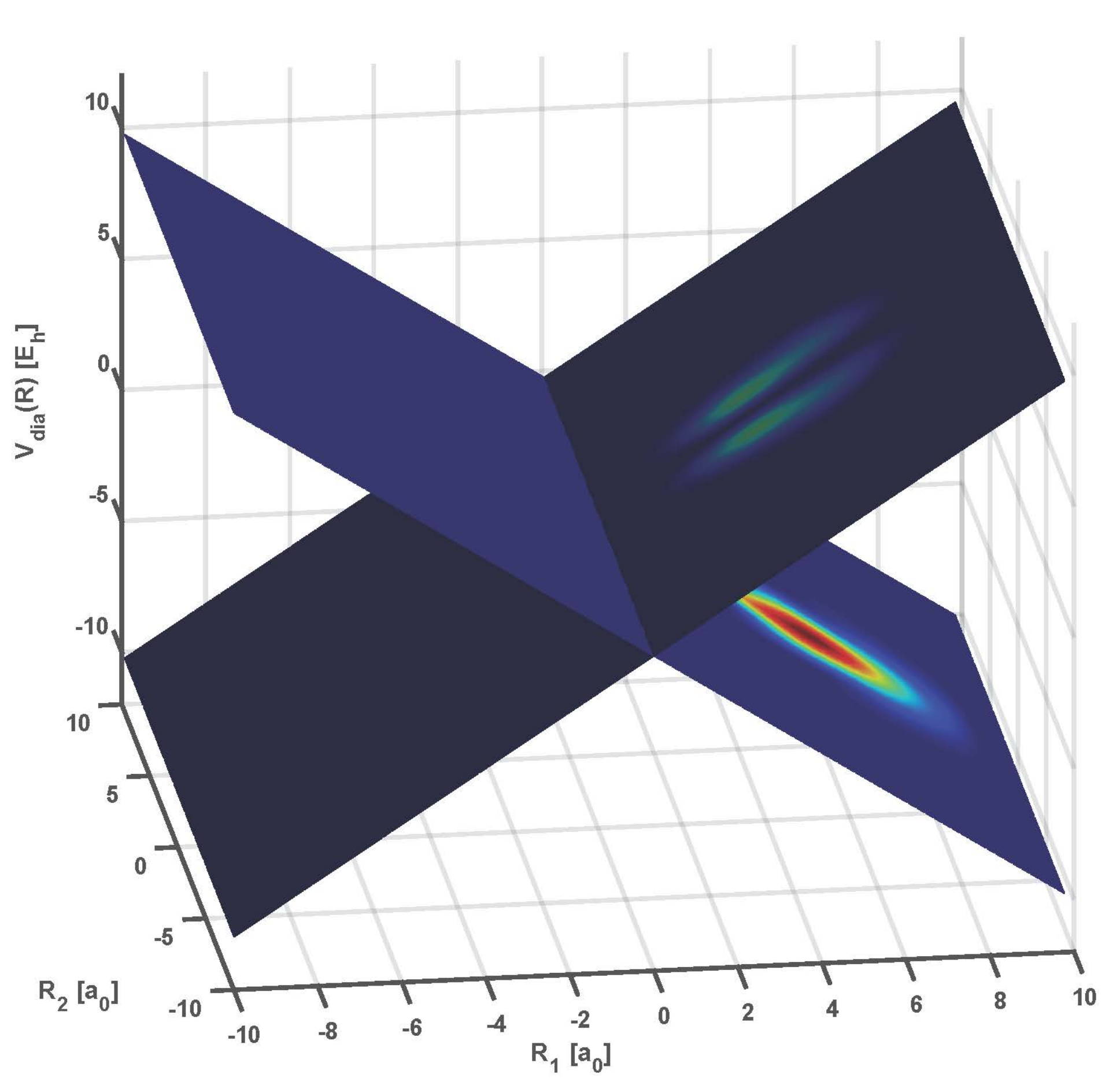}
		\includegraphics[width=10cm]{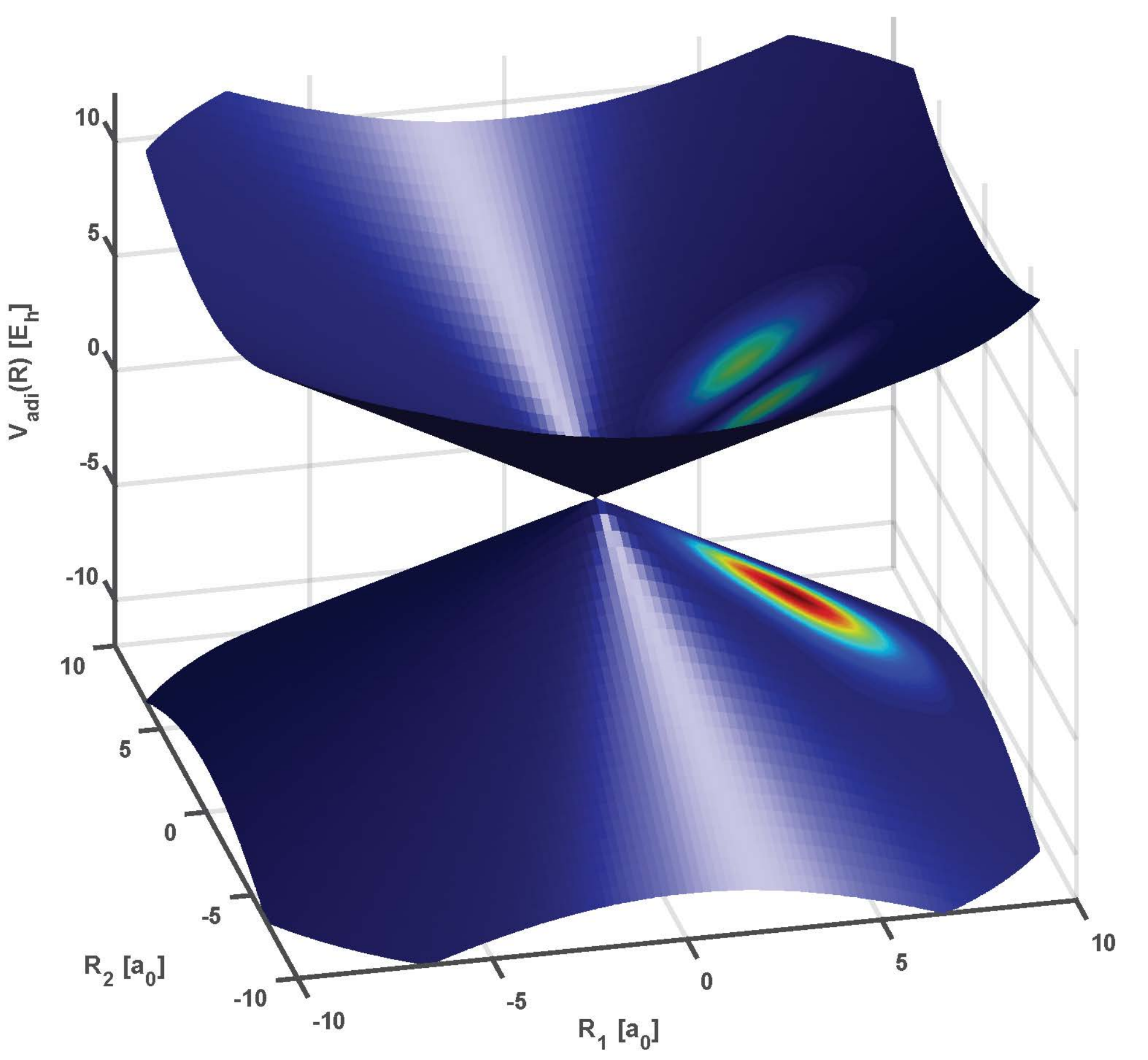}
	\label{fig:conical}
\end{figure}

Even though \textsc{WavePacket} uses internally the diabatic representation for coupled Schr\"odinger equations, in certain cases it may be advantageous to interpret quantum dynamics in an adiabatic representation.
There the potential energy matrix becomes diagonal, and non-adiabatic coupling tensors (NACTs) are introduced through a non-diagonal kinetic energy operator matrix. 
Hence, \textsc{WavePacket} offers a transformation of calculated wave function $\Psi(R,t)$, along with the potentials $\hat{V}(R)$, to the adiabatic picture as an option for the graphical output by setting the corresponding keyword \texttt{hamilt.coupling.coupling='adi'}. 
The prototypical example of Fig.~\ref{fig:conical} shows two linear potentials intersecting along a linear seam in a diabatic representation arising, e.~g., for a linear $E \otimes e$ Jahn-Teller system \cite{Worth:04a,Lasser:05a}.
Upon transformation to the adiabatic representation, the resulting potential energy surfaces form a conical intersection which is a prototypical system for non-adiabatic dynamics \cite{Truhlar:03a}, playing an important role e.~g., in photochemistry \cite{Baer:06a,Domcke:04a}.
This representation can give additional insights into the quantum dynamics in coupled channel calculations, especially when the adiabatic (Born-Oppenheimer or Born-Huang) approximation of vanishing NACTs is invoked \cite{Hagedorn2001,Panati2007,BSchmidt:39}.


\section{qm\_bound: Solving the time-independent Schr\"{o}dinger equation}
\label{sec::qm_bound}
The time-independent Schr\"{o}dinger equation (TISE) has the form 
\begin{equation}
\hat H_0(R,-i\nabla_R)\Psi_n(R)=E_n\Psi_n(R)
\label{eq::tise::eom}
\end{equation}
with eigenvalues $E_n$ and eigenfunctions $\Psi_n(R)$ of a time-independent Hamiltonian $\hat{H}_0$, i.~e., the Hamiltonian given in Eq.~(\ref{eq:Hamilt}) but without the last two terms on the right-hand side.
Note that for close coupling calculations, $\nu>1$, the operator $\hat{H}_0$ is an operator matrix and the wavefunctions $\Psi_n$ become matrix-valued. 

The expectation value of an observable $\hat O$ for the $n$-th eigenstate are obtained by the usual prescription
\begin{equation}
\langle \hat O \rangle_n = \langle\Psi_n|\hat O|\Psi_n\rangle .
\label{eq::tise::expect}
\end{equation}
Again, for $\nu>1$, \textsc{WavePacket} calculates these values both for each component of the operator $\hat O$ as well as for the complete quadratic form.
In \textsc{WavePacket}, the TISE (\ref{eq::tise::eom}) can be solved either directly using \texttt{qm\_bound} or indirectly by an imaginary time propagation using \texttt{qm\_propa}, as detailed in the following.

\subsection{Direct method}
\label{sec::qm_bound::direct}
Expanding the Hamiltonian and the eigenstates in one of the FBR basis sets introduced in Sec. \ref{sec::qm_init}, or the corresponding DVR grid representations, transforms the TISE~(\ref{eq::tise::eom}) into a conventional eigenvalue problem for the resulting matrix representation of the Hamiltonian (\ref{eq::dvr::fbr}).
This includes the widespread Fourier Grid Hamiltonian (FGH) method, which is equivalent to setting up the Hamiltonian matrix in a plane-wave DVR \cite{Meyer:70a,Marston:89a}, and which can be also generalized to the multidimensional case \cite{Webb:00a}.

Within \textsc{WavePacket}, the function \texttt{qm\_bound} sets up the Hamiltonian matrix and diagonalizes it using the \textsc{Matlab} command \texttt{eig} which first converts the matrix into a Hessenberg form and then determines the eigenvalues and eigenvectors by the QR method. 
Optionally, sparsity of the Hamiltonian matrix can be enforced by setting all entries below a user-specified threshold to zero.
This allows to exploit the advanced \textsc{Matlab} functionalities for dealing with sparse matrices, in particular diagonalization with \texttt{eigs} which is based on an Arnoldi algorithm. 
Note that methods avoiding an explicit setup of a Hamiltonian matrix, e.~g., Krylov subspace methods and variants thereof, are to be implemented in upcoming versions of \textsc{WavePacket}.
Moreover, note that \texttt{qm\_bound} is restricted to the calculation of bound states; the determination of scattering states is also planned for the near future.

The function \texttt{qm\_bound}  also offers limited support for symmetry-adaption of DVR schemes. 
Currently, symmetry adaption is still restricted to one-dimensional problems with a center of inversion where there is a choice between even and odd parity TISE solutions; further symmetries can be added upon future request of \textsc{WavePacket} users.
Note that symmetry adaption reduces not only the computational effort but can also increase the accuracy of bound state calculations.
An example are eigenstates in the 'deep tunneling' regime of a symmetric double well \cite{BSchmidt:67}, where a naive DVR approach would yield solutions that are not exact eigenstates of the parity operator, with the tunnel splitting deviating from the correct value.

\subsection{Relaxation method}
\label{sec::qm_bound::relax}
The direct diagonalization method to solve the TISE (\ref{eq::tise::eom}) as implemented in \texttt{qm\_bound} is simple, reliable and generally yields accurate results, even for highly excited states.
However, the matrices tend to get very large already for medium--dimensional problems, even if one exploits sparsity; in practice, direct diagonalization is currently limited to up to three-dimensional problems. 

As an alternative, \textsc{WavePacket} offers the ``relaxation method'' which is based on a propagation in imaginary time \cite{Kosloff:86a}.
Using an imaginary time variable $\tau = i t$, the TDSE (\ref{eq::tdse::eom})  transforms into a diffusion equation
\begin{equation}
	\frac{\partial}{\partial\tau} \Psi(R,\tau) = - \hat H_0(R,-i\nabla_R) \Psi(R,\tau),\quad\Psi(R,\tau=0)=\Psi_0(R)
\end{equation}
with the formal solution
\begin{equation}
\label{eq::tise::relaxation}
	\Psi(\tau) 
	= \mathrm{e}^{-\hat H_0 (\tau - \tau_0)} \Psi(\tau_0) = \sum_{n=0}^\infty \mathrm{e}^{-E_n (\tau - \tau_0)} \Psi_n \langle \Psi_n | \Psi(\tau_0)\rangle
	,
\end{equation}
for the eigenvectors and -values defined in Eq.~(\ref{eq::tise::eom}).
For sufficiently large $\tau$, the wave function is dominated by the ground state and states (nearly) degenerate with it.
By choosing the initial wave function $\Psi(\tau_0)$ appropriately, it is also possible to enforce the symmetry in the case of (near-)degenerate states.

Excited states are obtained by repeating the relaxation in the subspace orthogonal to the ground state, i.~e., by projecting out the ground state.
By iteratively projecting out more and more states, it is in principle possible to compute even higher excited states.
However, errors from finite convergence, i.~e., from a finite value of $\tau$, quickly accumulate over multiple iterations.
In practice, often only a handful of lowest energy eigenstates can be calculated accurately in this way.
This approach has not yet been implemented in the current version 5.2.1 of \textsc{WavePacket}; see, however, our demonstration example of vibrational states of the H$_3^+$ molecule for a prototype implementation.

In principle, every numeric scheme that solves the Schr\"odinger equation in real time can be adapted to work with imaginary time.
However, we are generally only interested in the long-time limit and time-independent Hamiltonians where again propagation schemes based on Chebychev polynomials  are particularily efficient \cite{Kosloff:86a}.
Hence, only the Chebychev method is available for imaginary time propagation using the \texttt{qm\_propa} function in \textsc{WavePacket}. 
It is essentially identical to the scheme described in Sec. \ref{sec::qm_propa::cheb} but with the expansion coefficients being proportional to modified Bessel functions of the first kind, $c_n(\alpha) \propto I_n(\alpha)$.
Finally, we remark that imaginary time propagation also occurs in some other contexts, such as random wave function methods that deal with finite temperatures \cite{Gelman:03a}.

\subsection{Example: Bound states of a Morse oscillator}

\begin{figure}
	\caption{Wigner representation of the highest bound state of a Morse oscillator obtained using \texttt{qm\_bound}. The marginals show the position and momentum densities}
	\includegraphics[width=12cm]{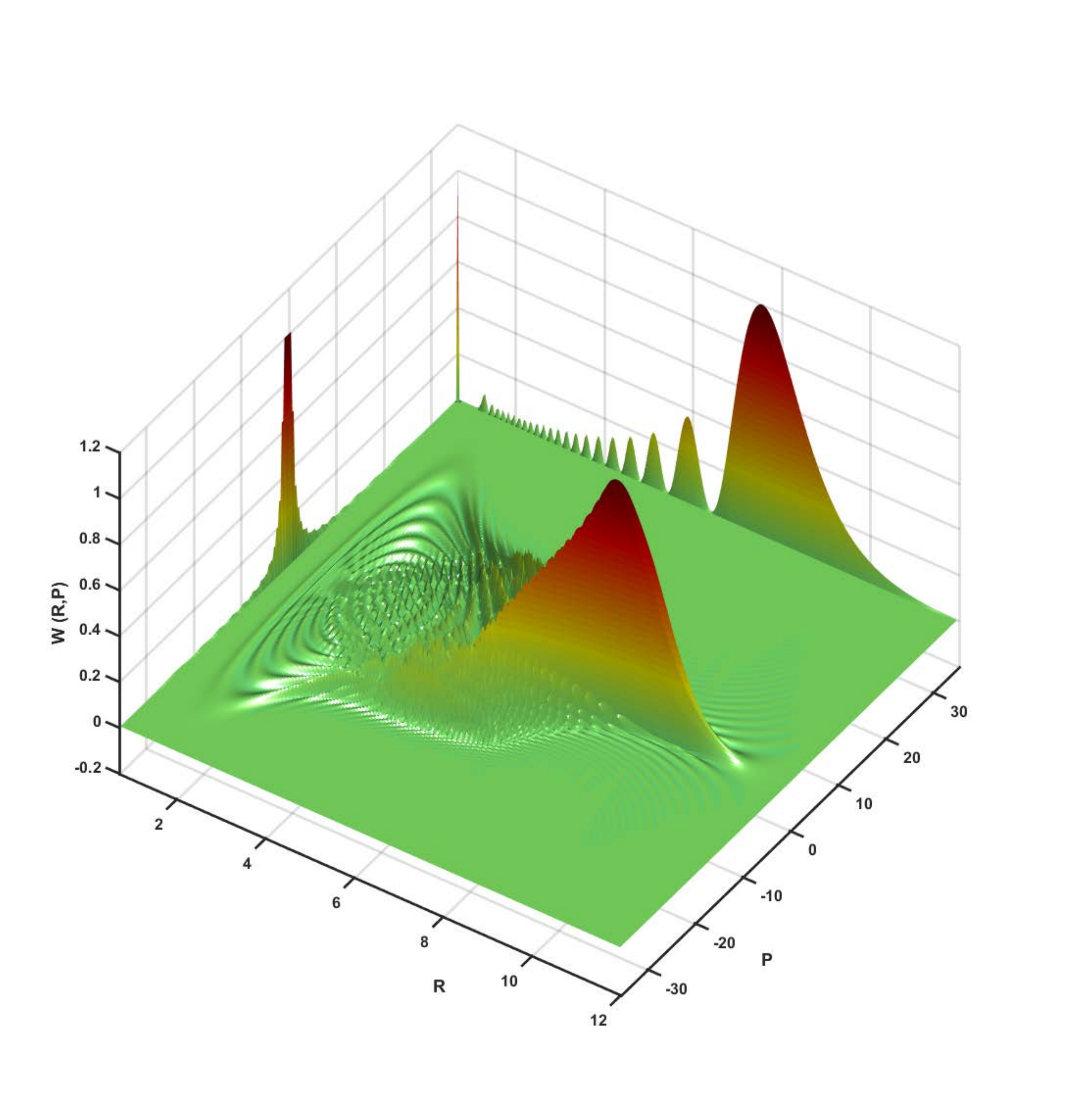}
	\label{fig:bound}
\end{figure}

Here we come back to the previous example of the Morse oscillator introduced in Eqs.~(\ref{eq::morse}), (\ref{eq::mecke}) in Sec.~\ref{sec::qm_init::morse}.
For a bound state calculation using \texttt{qm\_bound} with the FFT grid representation introduced in Sec.~\ref{sec::qm_init::fft} only one additional line (\texttt{psi.eigen.stop = 21;}) has to be added to \texttt{qm\_init} to calculate all vibrational states with $0\leq v \leq 21$.
Fig.~\ref{fig:bound} shows the highest vibrational state which is strongly anharmonic with $\langle R\rangle_{21} \approx 4 R_e$. Even though its energy $E_{21}$ is only 0.5 \% of the well depth $D_e$ below the dissociation limit, this energy is calculated with a relative error of only $10^{-5}$ when using 256 grid points.

As an alternative to the FFT-based DVR method, the Gauss-Hermite scheme can be used which requires the following settings
\begin{verbatim}
space.dof{1} = grid_hermite; 
space.dof{1}.mass = 1728.539; 
space.dof{1}.omega = 0.0172;
space.dof{1}.r_e = 1.821; 
space.dof{1}.n_pts = 25; 
\end{verbatim}
where the parameters are chosen such as to resemble the harmonic approximation of the Morse oscillator. 
Hence, for low excitation, this DVR scheme can be expected to be most efficient (here only 25 basis functions), whereas the FFT scheme will be usually preferred for the description of highly excitated vibrational or continuum states \cite{BSchmidt:14,BSchmidt:17,BSchmidt:20}. 

When using the relaxation method instead of the direct methods, the imaginary time propagator is expanded in Chebychev polynomials (\texttt{time.propa.handle=@ket.cheby\_imag;}) whereas the time stepping can remain the same as in the real time Chebychev propagation shown in Sec.~\ref{sec::qm_propa::rev}, here resulting in the use of 87 polynomials.
As an initial wavefunction one may choose a Gaussian packet
\begin{equation}
\Psi(R,t=0) \equiv \Psi_0 = \exp\left[-\left(\frac{x-x_0}{2w}\right)^2\right]
\end{equation}
centered around $x_0$ with uncertainty $w$. 
Such a choice can be realized by the following lines of \textsc{MATLAB} code
\begin{verbatim}
psi.init.dof{1}.handle= @wav.gauss; 
psi.init.dof{1}.pos_0 =  4.0;
psi.init.dof{1}.width =  0.5; 
\end{verbatim}
which calls the function \texttt{gauss} from the \texttt{+wav} package folder to set up the initial wave function.
With this choice of parameters, the wave function relaxes to the ground state already after 3 main time steps.

We want to remark that the above-mentioned procedure to obtain excited states is currently not implemented in \textsc{WavePacket}, although it can be scripted with moderate effort.
An example for the use of the projection to relax to higher excited states can be found among the demonstration examples (MolVibration/H3+).

\section{qm\_movie: Creating animated graphics}
\label{sec::qm_movie}

One of the key advantages of the \textsc{Matlab} version of \textsc{WavePacket} is its strong commitment to computer graphics.
This is not only mandatory for the use in teaching, but is also useful in research projects.
One of the hallmarks of \textsc{WavePacket} is its ability to create graphics automatically `on the fly', i.~e., one frame is created for each bound state during a TISE simulation with \texttt{qm\_bound} or for each main time step in a TDSE propagation using \texttt{qm\_propa}.
Thus, errors are discovered easily already while the simulation is still running, and the user can develop a more intuitive understanding of the quantum dynamics.

In addition to the generation of graphical output on the fly, we have also added a function \texttt{qm\_movie} that produces animated graphics from saved wave functions.
As a first step, it is necessary to have saved the wavefunctions while running \texttt{qm\_bound} or \texttt{qm\_propa}, which is done by setting the following keyword in \texttt{qm\_init}
\begin{verbatim}
psi.save.export = true;
\end{verbatim}
By default, the wavefunctions are stored in the current working directory; alternative choices are possible by setting \texttt{psi.save.dir} appropriately.
All global variables containing the general simulation settings are stored in a file usually named \texttt{WavePacketSave.mat} (other choices to be specified through \texttt{psi.save.file}), while the wavefunctions themselves are stored in files \texttt{WavePacketSave}\_\textit{n}.\texttt{mat}.
The splitting into several files, $n=0,1,\ldots$, is intended to balance the requirements of memory size and execution speed.

\textsc{WavePacket} comprises a variety of different graphical representations of the probability density $\rho(R)=|\Psi(R)|^2$.
The appearance of the plots can be controlled by the variable \texttt{plots.density.type}, which is typically also set in \texttt{qm\_init}.
Depending on the dimensionality of the system, this variable can take the values listed below, which mostly correspond to functions inside the package folder \texttt{+plot}.
The animated graphical output is normally saved in the working directory, as an MP4 files with name corresponding to \texttt{plots.density.type} and with extension \texttt{.mpg}.

\subsection{Wavepackets in one dimension}
\begin{description}
	\item['curve':] 
    Curve plot of the density in position (DVR) or in momentum (FBR) representation, with color coding of the complex phase of the wavefunction $\Psi(R)$
   \item ['polar':]
    Same as 'curve', but using a polar representation 
\item ['contour':]
    Contour plot of Wigner quasi-probability distribution~\cite{Hillery:84a}
		\begin{equation}
		W(R,P) \propto \int dY \exp(iPY) \Psi^*(R+Y/2) \Psi(R-Y/2)
		\end{equation}
		optionally with position and momentum densities as marginals (FFT grid only). For an example, see Fig.~\ref{fig:revival}.
\item ['surface':]
    Surface plot of Wigner distribution, see above (FFT grid only). For an example, see Fig.~\ref{fig:bound}.
\item ['flux':]
    Curve plot of the probability flux
		\begin{equation}
		J(R)=\Re\left(-i\Psi^*(R)\nabla_R\Psi(R)\right)
		\end{equation}
\end{description}

\subsection{Wavepackets in two dimensions}
\begin{description}
	\item ['contour':]
    Contour plot of density in position (DVR) or in momentum (FBR) representation
\item ['surface':]
    Surface plot of density in position (DVR) or in momentum (FBR) representation. For an example, see Fig.~\ref{fig:conical}.
\item ['flux':]
    Quiver plots of probability flux vectors in position (DVR) representation
\end{description}

\subsection{Wavepackets in three dimensions}
\begin{description}
	\item ['surface':]
    Iso-surface in position (DVR) or in momentum (FBR) representation
\end{description}

\subsection {For any number of dimensions}
\begin{description}
	\item ['reduced':]
    Reduced densities, $\bar\rho_k$, in each dimension $k$, obtained by tracing over all other degrees of freedom, see e.~g. Ref.~\cite{BSchmidt:62}, shown along with corresponding purity measures, $\mathrm{tr}(\bar\rho_k^2)$
\end{description}

\subsection{Further settings}
In addition to the choice of the type of graphical representation, there are many further parameters governing the details of the graphics output.
As a few important examples we mention here the following keywords
\begin{itemize}
\item
\texttt{plots.density.representation} has to be set to \texttt{'fbr'} or \texttt{'dvr'} to switch between FBR and DVR, respectively, for some of the graphical representations listed above
\item 
\texttt{plots.density.energy.on} is used to toggle the representation of potential/kinetic/total energy curves/surfaces to be displayed along with the densities. 
Their minimal/maximal values may be set through \texttt{plots.expect.energies.min|max}
\item
\texttt{plots.expect.on} is used to toggle curve plots of various expectation values, optionally with error bars (\texttt{plots.expect.errorbar.on=true}) indicating the respective quantum-mechanical uncertainties. Also shown is the autocorrelation function of the wavefunction
\item
\texttt{plots.spectrum.on} is used to toggle a curve plot of a spectrum obtained as a Fourier transform of the autocorrelation function 
\end{itemize}
Default settings for these and many other parameters can be found inside the function \texttt{+init/plots} which is called by function \texttt{qm\_setup}, see Sec. \ref{sec::qm_setup}.
They can be arbitrarily overwritten by the user in a subsequent call to \texttt{qm\_init} before being consumed in \texttt{qm\_propa}, \texttt{qm\_bound}, or \texttt{qm\_movie}.

\section{Summary and outlook}

In the present Part I, we have introduced the functions \texttt{qm\_propa} and \texttt{qm\_bound} as the main tools to numerically solve the TDSE and the TISE, respectively, within the framework of the \textsc{WavePacket} software package (version 5.2.1). 
They are accompanied by the function \texttt{qm\_movie} to produce animated graphics, as well as the auxiliary functions \texttt{qm\_setup} and \texttt{qm\_cleanup} to initialize and finalize simulation protocols.
All the user-defined settings should be typically combined in a function \texttt{qm\_init} for which sample code lines and discussions can be found scattered throughout this work.
Although all of these explanations were only for the example of a simple Morse oscillator, generalizations to other and more complex simulation tasks using \textsc{WavePacket} should be straight-forward.
In principle, our software can handle an arbitrary number of degrees of freedom as well as an arbitrary number of coupled Schr\"{o}dinger equations. 
However, in practice, hardware limitations impose an upper limit to the size of the problems that can be treated with \textsc{WavePacket}. 
The following guidelines give you an idea of what can be achieved using a standard PC (equipped with 8 GB of memory) when using the FFT-based representations
\begin{itemize}
\item
For dynamical simulations using \texttt{qm\_propa}, the number of grid points in a simulation should typically not exceed $2^{24} \approx  1.677 \times 10^7$. 
Hence, three-dimensional simulations on a grid of $256^3$ or four-dimensional simulations on a grid of $64^4$ points should roughly give the upper limit.
\item
For bound state calculations using \texttt{qm\_bound}, the use of large matrices for the Fourier grid Hamiltonian (FGH) algorithm reduces the maximal grid sizes roughly to the square root of the above numbers, i.e., to a total of $2^{12}$ grid points. 
However, by virtue of \textsc{MATLAB}'s support for sparse matrices, grid sizes up to $2^{15...16} = 32768 \ldots 65536$ appear possible. 
This is sufficient for a two- or three-dimensional simulation employing $256^2$ or $32^3$ grid points, respectively.
However, when using imaginary time propagation (see Sec.~\ref{sec::qm_bound::relax}), the same number of DVR points as for \texttt{qm\_propa} is possible.
\end{itemize}
If quantum dynamics simulations in more dimensions are required, one may consider using the MCTDH program package \cite{Beck2000} which is based on multi-configurational time-dependent Hartree algorithm \cite{Meyer:90a,Manthe:92a}. 
MCTDH will in general be best suited for systems with 4 to 12 degrees of freedom; for certain classes of problems it is possible to treat even much larger systems. 
While \textsc{WavePacket} certainly cannot compete with the MCTDH package with respect to high dimensionality, we emphasize, however, the intuitive handling of \textsc{WavePacket}, in particular its superior graphical capabilities and easy customizability.

Note that the present Part I is mainly centered on codes to solve Schr\"{o}dinger equations which were already present in previous \textsc{WavePacket} versions 4.x dating from a few years ago and which are already in use in several laboratories worldwide.  
The upcoming Part II will deal with more recent parts of \textsc{WavePacket}, mainly introduced in versions 5.x released since 2014.
Among others there are extensions to treat also open quantum systems by solving Liouville--von Neumann equations (LvNEs), with Lindblad operators to model dissipation and dephasing \cite{BSchmidt:61}.
For the case of quantum  systems driven by external fields, the latest \textsc{WavePacket} versions also contain different techniques for dimension reduction which are especially important when dealing with large scale LvNEs.
Other extensions of \textsc{WavePacket} are concerned with optimal control of quantum dynamics \cite{Ohtsuki2004}, introduced in version 5.2.0.

While the oldest versions of \textsc{WavePacket} were written in a relatively traditional, completely procedural way, the introduction of generalized DVR methods in version 4.5 required a significant restructuring of large parts of the code, which led us to make use of some of the object-oriented features offered by \textsc{Matlab}. 
This was instrumental in making the codes more flexible to cover the growing diversity of physical/chemical systems simulated with \textsc{WavePacket}, in particular allowing for an easier implementation of different types of kinetic energy operators.
However, there still remain issues which are difficult to implement in the \textsc{Matlab} version of \textsc{WavePacket}.
In particular, it is difficult to treat propagations of quantum states and of density operators on an equal footing.
Moreover, it is not easy to propagate a system not only quantum-mechanically but also classically \cite{BSchmidt:38} or using mixed quantum-classical dynamics \cite{BSchmidt:39,BSchmidt:43}, although the mathematics of the DVR/FBR expansions are rather similar.
To overcome these inherent limitations, we have recently set up a C++ project  at \textsc{SourceForge} aiming at a rewrite of the \textsc{WavePacket} codes in a completely object-oriented manner.
However, with this project still being in an early stage of its development, the reader can expect the \textsc{Matlab} codes presented here to remain the main working version of \textsc{WavePacket} over the next few years.

\begin{acknowledgments}
This work has been supported by the Einstein Center for Mathematics Berlin (ECMath) through project SE~11 and by the Deutsche Forschungsgemeinschaft (DFG) through grants SCHM 1202/3-1 and Sa 547/9.
Furthermore, fruitful discussions with C.~Hartmann, J.~Manz, P.~Saalfrank, Ch.~Salzmann, S.~Schulz, and Ch.~Sch\"{u}tte at Free University Berlin are gratefully acknowledged.
\end{acknowledgments}

\bibliography{WavePacket1}

\begin{thebibliography}{67}
\expandafter\ifx\csname natexlab\endcsname\relax\def\natexlab#1{#1}\fi
\expandafter\ifx\csname bibnamefont\endcsname\relax
  \def\bibnamefont#1{#1}\fi
\expandafter\ifx\csname bibfnamefont\endcsname\relax
  \def\bibfnamefont#1{#1}\fi
\expandafter\ifx\csname citenamefont\endcsname\relax
  \def\citenamefont#1{#1}\fi
\expandafter\ifx\csname url\endcsname\relax
  \def\url#1{\texttt{#1}}\fi
\expandafter\ifx\csname urlprefix\endcsname\relax\def\urlprefix{URL }\fi
\providecommand{\bibinfo}[2]{#2}
\providecommand{\eprint}[2][]{\url{#2}}

\bibitem[{\citenamefont{May and K\"{u}hn}(2000)}]{May:00a}
\bibinfo{author}{\bibfnamefont{V.}~\bibnamefont{May}} \bibnamefont{and}
  \bibinfo{author}{\bibfnamefont{O.}~\bibnamefont{K\"{u}hn}},
  \emph{\bibinfo{title}{{Charge and Energy Transfer Dynamics in Molecular
  Systems}}} (\bibinfo{publisher}{Wiley}, \bibinfo{address}{Berlin},
  \bibinfo{year}{2000}).

\bibitem[{\citenamefont{Schleich}(2001)}]{Schleich:01a}
\bibinfo{author}{\bibfnamefont{W.~P.} \bibnamefont{Schleich}},
  \emph{\bibinfo{title}{{Quantum Optics in Phase Space}}}
  (\bibinfo{publisher}{Wiley--VCH}, \bibinfo{address}{Berlin},
  \bibinfo{year}{2001}).

\bibitem[{\citenamefont{Tannor}(2004)}]{Tannor:04a}
\bibinfo{author}{\bibfnamefont{D.}~\bibnamefont{Tannor}},
  \emph{\bibinfo{title}{{Introduction to Quantum Mechanics. A Time-Dependent
  Perspective}}} (\bibinfo{publisher}{University Science Books},
  \bibinfo{address}{Sausalito}, \bibinfo{year}{2004}).

\bibitem[{\citenamefont{Gro{\ss}mann}(2008)}]{Grossmann2008}
\bibinfo{author}{\bibfnamefont{F.}~\bibnamefont{Gro{\ss}mann}},
  \emph{\bibinfo{title}{{Theoretical Femtosecond Physics}}}
  (\bibinfo{publisher}{Springer}, \bibinfo{address}{Berlin Heidelberg},
  \bibinfo{year}{2008}).

\bibitem[{\citenamefont{Zewail}(2000)}]{Zewail2000}
\bibinfo{author}{\bibfnamefont{A.~H.} \bibnamefont{Zewail}},
  \bibinfo{journal}{J. Phys. Chem. A} \textbf{\bibinfo{volume}{104}},
  \bibinfo{pages}{5660} (\bibinfo{year}{2000}).

\bibitem[{\citenamefont{Sundstr{\"{o}}m}(2008)}]{Sundstrom2008}
\bibinfo{author}{\bibfnamefont{V.}~\bibnamefont{Sundstr{\"{o}}m}},
  \bibinfo{journal}{Ann. Rev. Phys. Chem.} \textbf{\bibinfo{volume}{59}},
  \bibinfo{pages}{53} (\bibinfo{year}{2008}).

\bibitem[{\citenamefont{de~Vivie-Riedle and Troppmann}(2007)}]{Vivie:07a}
\bibinfo{author}{\bibfnamefont{R.}~\bibnamefont{de~Vivie-Riedle}}
  \bibnamefont{and}
  \bibinfo{author}{\bibfnamefont{U.}~\bibnamefont{Troppmann}},
  \bibinfo{journal}{Chem. Rev.} \textbf{\bibinfo{volume}{107}},
  \bibinfo{pages}{5082} (\bibinfo{year}{2007}).

\bibitem[{\citenamefont{Kais}(2014)}]{Kais2014}
\bibinfo{editor}{\bibfnamefont{S.}~\bibnamefont{Kais}}, ed.,
  \emph{\bibinfo{title}{{Quantum Information and Computation for Chemistry}}}
  (\bibinfo{publisher}{Wiley}, \bibinfo{year}{2014}).

\bibitem[{\citenamefont{Beck et~al.}(2000)\citenamefont{Beck, J\"{a}ckle,
  Worth, and Meyer}}]{Beck2000}
\bibinfo{author}{\bibfnamefont{M.~H.} \bibnamefont{Beck}},
  \bibinfo{author}{\bibfnamefont{A.}~\bibnamefont{J\"{a}ckle}},
  \bibinfo{author}{\bibfnamefont{G.~A.} \bibnamefont{Worth}}, \bibnamefont{and}
  \bibinfo{author}{\bibfnamefont{H.-D.} \bibnamefont{Meyer}},
  \bibinfo{journal}{Phys. Rep.} \textbf{\bibinfo{volume}{324}},
  \bibinfo{pages}{1} (\bibinfo{year}{2000}).

\bibitem[{\citenamefont{Light et~al.}(1985)\citenamefont{Light, Hamilton, and
  Lill}}]{Light:85a}
\bibinfo{author}{\bibfnamefont{J.~C.} \bibnamefont{Light}},
  \bibinfo{author}{\bibfnamefont{I.~P.} \bibnamefont{Hamilton}},
  \bibnamefont{and} \bibinfo{author}{\bibfnamefont{J.~V.} \bibnamefont{Lill}},
  \bibinfo{journal}{J. Chem. Phys.} \textbf{\bibinfo{volume}{82}},
  \bibinfo{pages}{1400} (\bibinfo{year}{1985}).

\bibitem[{\citenamefont{Light and Carrington}(2000)}]{Light:00a}
\bibinfo{author}{\bibfnamefont{J.~C.} \bibnamefont{Light}} \bibnamefont{and}
  \bibinfo{author}{\bibfnamefont{T.}~\bibnamefont{Carrington}},
  \bibinfo{journal}{Adv. Chem. Phys.} \textbf{\bibinfo{volume}{114}},
  \bibinfo{pages}{263} (\bibinfo{year}{2000}).

\bibitem[{\citenamefont{Leforestier et~al.}(1991)\citenamefont{Leforestier,
  Bisseling, Cerjan, Feit, Friesner, Guldberg, Hammerich, Jolicard, Karrlein,
  Meyer et~al.}}]{Leforestier:91a}
\bibinfo{author}{\bibfnamefont{C.}~\bibnamefont{Leforestier}},
  \bibinfo{author}{\bibfnamefont{R.}~\bibnamefont{Bisseling}},
  \bibinfo{author}{\bibfnamefont{C.}~\bibnamefont{Cerjan}},
  \bibinfo{author}{\bibfnamefont{M.}~\bibnamefont{Feit}},
  \bibinfo{author}{\bibfnamefont{R.}~\bibnamefont{Friesner}},
  \bibinfo{author}{\bibfnamefont{A.}~\bibnamefont{Guldberg}},
  \bibinfo{author}{\bibfnamefont{A.}~\bibnamefont{Hammerich}},
  \bibinfo{author}{\bibfnamefont{G.}~\bibnamefont{Jolicard}},
  \bibinfo{author}{\bibfnamefont{W.}~\bibnamefont{Karrlein}},
  \bibinfo{author}{\bibfnamefont{H.-D.} \bibnamefont{Meyer}},
  \bibnamefont{et~al.}, \bibinfo{journal}{J. Comp. Phys.}
  \textbf{\bibinfo{volume}{94}}, \bibinfo{pages}{59} (\bibinfo{year}{1991}).

\bibitem[{\citenamefont{Kan et~al.}(2013)\citenamefont{Kan, Sardar, Sahoo,
  Sarkar, and Adhikari}}]{Khan2013}
\bibinfo{author}{\bibfnamefont{B.~A.} \bibnamefont{Kan}},
  \bibinfo{author}{\bibfnamefont{S.}~\bibnamefont{Sardar}},
  \bibinfo{author}{\bibfnamefont{T.}~\bibnamefont{Sahoo}},
  \bibinfo{author}{\bibfnamefont{P.}~\bibnamefont{Sarkar}}, \bibnamefont{and}
  \bibinfo{author}{\bibfnamefont{S.}~\bibnamefont{Adhikari}},
  \bibinfo{journal}{J. Theo. Comp. Chem.} \textbf{\bibinfo{volume}{12}},
  \bibinfo{pages}{1350042} (\bibinfo{year}{2013}).

\bibitem[{\citenamefont{Baer}(2006)}]{Baer:06a}
\bibinfo{author}{\bibfnamefont{M.}~\bibnamefont{Baer}},
  \emph{\bibinfo{title}{{Beyond Born--Oppenheimer}}}
  (\bibinfo{publisher}{Wiley-VCH}, \bibinfo{address}{Hoboken, New Jersey},
  \bibinfo{year}{2006}).

\bibitem[{\citenamefont{Domcke et~al.}(2004)\citenamefont{Domcke, Yarkony, and
  K\"{o}ppel}}]{Domcke:04a}
\bibinfo{editor}{\bibfnamefont{W.}~\bibnamefont{Domcke}},
  \bibinfo{editor}{\bibfnamefont{D.~R.} \bibnamefont{Yarkony}},
  \bibnamefont{and}
  \bibinfo{editor}{\bibfnamefont{H.}~\bibnamefont{K\"{o}ppel}}, eds.,
  \emph{\bibinfo{title}{{Conical Intersections. Electronic Structure, Dynamics
  and Spectroscopy}}}, vol.~\bibinfo{volume}{15} of
  \emph{\bibinfo{series}{Advanced Series in Physical Chemistry}}
  (\bibinfo{publisher}{World Scientific}, \bibinfo{address}{Singapore},
  \bibinfo{year}{2004}).

\bibitem[{\citenamefont{Weiss}(1999)}]{Weiss:99a}
\bibinfo{author}{\bibfnamefont{U.}~\bibnamefont{Weiss}},
  \emph{\bibinfo{title}{{Quantum Dissipative Systems}}}
  (\bibinfo{publisher}{World Scientific}, \bibinfo{address}{Singapore},
  \bibinfo{year}{1999}).

\bibitem[{\citenamefont{Breuer and Petruccione}(2002)}]{Breuer:02a}
\bibinfo{author}{\bibfnamefont{H.-P.} \bibnamefont{Breuer}} \bibnamefont{and}
  \bibinfo{author}{\bibfnamefont{F.}~\bibnamefont{Petruccione}},
  \emph{\bibinfo{title}{{The Theory of Open Quantum Systems}}}
  (\bibinfo{publisher}{Oxford University Press}, \bibinfo{year}{2002}).

\bibitem[{\citenamefont{Mecke}(1950)}]{Mecke:50a}
\bibinfo{author}{\bibfnamefont{R.}~\bibnamefont{Mecke}}, \bibinfo{journal}{Z.
  Elektrochem.} \textbf{\bibinfo{volume}{54}}, \bibinfo{pages}{38}
  (\bibinfo{year}{1950}).

\bibitem[{\citenamefont{Korolkov
  et~al.}(1996{\natexlab{a}})\citenamefont{Korolkov, Paramonov, and
  Schmidt}}]{BSchmidt:14}
\bibinfo{author}{\bibfnamefont{M.~V.} \bibnamefont{Korolkov}},
  \bibinfo{author}{\bibfnamefont{G.~K.} \bibnamefont{Paramonov}},
  \bibnamefont{and} \bibinfo{author}{\bibfnamefont{B.}~\bibnamefont{Schmidt}},
  \bibinfo{journal}{J. Chem. Phys.} \textbf{\bibinfo{volume}{105}},
  \bibinfo{pages}{1862} (\bibinfo{year}{1996}{\natexlab{a}}).

\bibitem[{\citenamefont{Kosloff and Kosloff}(1983)}]{Kosloff:83a}
\bibinfo{author}{\bibfnamefont{D.}~\bibnamefont{Kosloff}} \bibnamefont{and}
  \bibinfo{author}{\bibfnamefont{R.}~\bibnamefont{Kosloff}},
  \bibinfo{journal}{J Comp. Phys.} \textbf{\bibinfo{volume}{52}},
  \bibinfo{pages}{35} (\bibinfo{year}{1983}).

\bibitem[{\citenamefont{Fischer et~al.}(2011)\citenamefont{Fischer, Lorenz,
  Schmidt, and Schmidt}}]{BSchmidt:59}
\bibinfo{author}{\bibfnamefont{M.}~\bibnamefont{Fischer}},
  \bibinfo{author}{\bibfnamefont{U.}~\bibnamefont{Lorenz}},
  \bibinfo{author}{\bibfnamefont{B.}~\bibnamefont{Schmidt}}, \bibnamefont{and}
  \bibinfo{author}{\bibfnamefont{R.}~\bibnamefont{Schmidt}},
  \bibinfo{journal}{Phys. Rev. A} \textbf{\bibinfo{volume}{84}},
  \bibinfo{pages}{033422} (\bibinfo{year}{2011}).

\bibitem[{\citenamefont{Backhaus and Schmidt}(1997)}]{BSchmidt:19}
\bibinfo{author}{\bibfnamefont{P.}~\bibnamefont{Backhaus}} \bibnamefont{and}
  \bibinfo{author}{\bibfnamefont{B.}~\bibnamefont{Schmidt}},
  \bibinfo{journal}{Chem. Phys.} \textbf{\bibinfo{volume}{217}},
  \bibinfo{pages}{131} (\bibinfo{year}{1997}).

\bibitem[{\citenamefont{Korolkov and Schmidt}(1997)}]{BSchmidt:20}
\bibinfo{author}{\bibfnamefont{M.~V.} \bibnamefont{Korolkov}} \bibnamefont{and}
  \bibinfo{author}{\bibfnamefont{B.}~\bibnamefont{Schmidt}},
  \bibinfo{journal}{Chem. Phys. Lett.} \textbf{\bibinfo{volume}{272}},
  \bibinfo{pages}{96} (\bibinfo{year}{1997}).

\bibitem[{\citenamefont{Korolkov and Schmidt}(1998)}]{BSchmidt:26}
\bibinfo{author}{\bibfnamefont{M.~V.} \bibnamefont{Korolkov}} \bibnamefont{and}
  \bibinfo{author}{\bibfnamefont{B.}~\bibnamefont{Schmidt}},
  \bibinfo{journal}{Chem. Phys.} \textbf{\bibinfo{volume}{237}},
  \bibinfo{pages}{123} (\bibinfo{year}{1998}).

\bibitem[{\citenamefont{Korolkov and Schmidt}(2002)}]{BSchmidt:37}
\bibinfo{author}{\bibfnamefont{M.~V.} \bibnamefont{Korolkov}} \bibnamefont{and}
  \bibinfo{author}{\bibfnamefont{B.}~\bibnamefont{Schmidt}},
  \bibinfo{journal}{Chem. Phys. Lett.} \textbf{\bibinfo{volume}{361}},
  \bibinfo{pages}{432} (\bibinfo{year}{2002}).

\bibitem[{\citenamefont{Ring et~al.}(1998)\citenamefont{Ring, Schmidt, and
  Baumg\"{a}rtel}}]{BSchmidt:23}
\bibinfo{author}{\bibfnamefont{S.}~\bibnamefont{Ring}},
  \bibinfo{author}{\bibfnamefont{B.}~\bibnamefont{Schmidt}}, \bibnamefont{and}
  \bibinfo{author}{\bibfnamefont{H.}~\bibnamefont{Baumg\"{a}rtel}},
  \bibinfo{journal}{Eur. Phys. J. D} \textbf{\bibinfo{volume}{3}},
  \bibinfo{pages}{155} (\bibinfo{year}{1998}).

\bibitem[{\citenamefont{Ronen et~al.}(2004)\citenamefont{Ronen,
  Nachtigallov\'{a}, Schmidt, and Jungwirth}}]{BSchmidt:44}
\bibinfo{author}{\bibfnamefont{S.}~\bibnamefont{Ronen}},
  \bibinfo{author}{\bibfnamefont{D.}~\bibnamefont{Nachtigallov\'{a}}},
  \bibinfo{author}{\bibfnamefont{B.}~\bibnamefont{Schmidt}}, \bibnamefont{and}
  \bibinfo{author}{\bibfnamefont{P.}~\bibnamefont{Jungwirth}},
  \bibinfo{journal}{Phys. Rev. Lett.} \textbf{\bibinfo{volume}{93}},
  \bibinfo{pages}{48301} (\bibinfo{year}{2004}).

\bibitem[{\citenamefont{Manolopoulos and Wyatt}(1988)}]{Manolopoulos:88a}
\bibinfo{author}{\bibfnamefont{D.}~\bibnamefont{Manolopoulos}}
  \bibnamefont{and} \bibinfo{author}{\bibfnamefont{R.}~\bibnamefont{Wyatt}},
  \bibinfo{journal}{Chem. Phys. Lett.} \textbf{\bibinfo{volume}{152}},
  \bibinfo{pages}{23} (\bibinfo{year}{1988}).

\bibitem[{\citenamefont{Muckerman}(1990)}]{Muckerman:90a}
\bibinfo{author}{\bibfnamefont{J.~T.} \bibnamefont{Muckerman}},
  \bibinfo{journal}{Chem. Phys. Lett.} \textbf{\bibinfo{volume}{173}},
  \bibinfo{pages}{200} (\bibinfo{year}{1990}).

\bibitem[{\citenamefont{Askar and Cakmak}(1978)}]{Askar:78a}
\bibinfo{author}{\bibfnamefont{A.}~\bibnamefont{Askar}} \bibnamefont{and}
  \bibinfo{author}{\bibfnamefont{A.~S.} \bibnamefont{Cakmak}},
  \bibinfo{journal}{J. Chem. Phys.} \textbf{\bibinfo{volume}{68}},
  \bibinfo{pages}{2794} (\bibinfo{year}{1978}).

\bibitem[{\citenamefont{Fleck et~al.}(1976)\citenamefont{Fleck, Morris, and
  Feit}}]{Fleck:76a}
\bibinfo{author}{\bibfnamefont{J.~A.} \bibnamefont{Fleck}},
  \bibinfo{author}{\bibfnamefont{J.~R.} \bibnamefont{Morris}},
  \bibnamefont{and} \bibinfo{author}{\bibfnamefont{M.~D.} \bibnamefont{Feit}},
  \bibinfo{journal}{Appl. Phys.} \textbf{\bibinfo{volume}{10}},
  \bibinfo{pages}{129} (\bibinfo{year}{1976}).

\bibitem[{\citenamefont{Feit et~al.}(1982)\citenamefont{Feit, {Fleck, Jr}, and
  Steiger}}]{Feit:82a}
\bibinfo{author}{\bibfnamefont{M.}~\bibnamefont{Feit}},
  \bibinfo{author}{\bibfnamefont{J.}~\bibnamefont{{Fleck, Jr}}},
  \bibnamefont{and} \bibinfo{author}{\bibfnamefont{A.}~\bibnamefont{Steiger}},
  \bibinfo{journal}{J. Comp. Phys.} \textbf{\bibinfo{volume}{47}},
  \bibinfo{pages}{412} (\bibinfo{year}{1982}).

\bibitem[{\citenamefont{Bandrauk and Shen}(1992)}]{Bandrauk:92a}
\bibinfo{author}{\bibfnamefont{A.~D.} \bibnamefont{Bandrauk}} \bibnamefont{and}
  \bibinfo{author}{\bibfnamefont{H.}~\bibnamefont{Shen}},
  \bibinfo{journal}{Can. J. Chem.} \textbf{\bibinfo{volume}{70}},
  \bibinfo{pages}{555} (\bibinfo{year}{1992}).

\bibitem[{\citenamefont{Tal-Ezer and Kosloff}(1984)}]{Tal-Ezer:84a}
\bibinfo{author}{\bibfnamefont{H.}~\bibnamefont{Tal-Ezer}} \bibnamefont{and}
  \bibinfo{author}{\bibfnamefont{R.}~\bibnamefont{Kosloff}},
  \bibinfo{journal}{J. Chem. Phys.} \textbf{\bibinfo{volume}{81}},
  \bibinfo{pages}{3967} (\bibinfo{year}{1984}).

\bibitem[{\citenamefont{Chen and Guo}(1999)}]{Chen:99b}
\bibinfo{author}{\bibfnamefont{R.}~\bibnamefont{Chen}} \bibnamefont{and}
  \bibinfo{author}{\bibfnamefont{H.}~\bibnamefont{Guo}},
  \bibinfo{journal}{Comp. Phys. Comm.} \textbf{\bibinfo{volume}{119}},
  \bibinfo{pages}{19} (\bibinfo{year}{1999}).

\bibitem[{\citenamefont{Kosloff and Tal-Ezer}(1986)}]{Kosloff:86a}
\bibinfo{author}{\bibfnamefont{R.}~\bibnamefont{Kosloff}} \bibnamefont{and}
  \bibinfo{author}{\bibfnamefont{H.}~\bibnamefont{Tal-Ezer}},
  \bibinfo{journal}{Chem. Phys. Lett.} \textbf{\bibinfo{volume}{127}},
  \bibinfo{pages}{223} (\bibinfo{year}{1986}).

\bibitem[{\citenamefont{Huisinga et~al.}(1999)\citenamefont{Huisinga, Pesce,
  Kosloff, and Saalfrank}}]{Huisinga:99a}
\bibinfo{author}{\bibfnamefont{W.}~\bibnamefont{Huisinga}},
  \bibinfo{author}{\bibfnamefont{L.}~\bibnamefont{Pesce}},
  \bibinfo{author}{\bibfnamefont{R.}~\bibnamefont{Kosloff}}, \bibnamefont{and}
  \bibinfo{author}{\bibfnamefont{P.}~\bibnamefont{Saalfrank}},
  \bibinfo{journal}{J. Chem. Phys.} \textbf{\bibinfo{volume}{110}},
  \bibinfo{pages}{5538} (\bibinfo{year}{1999}).

\bibitem[{\citenamefont{Glauber}(1963)}]{Glauber:63a}
\bibinfo{author}{\bibfnamefont{R.}~\bibnamefont{Glauber}},
  \bibinfo{journal}{Phys. Rev.} \textbf{\bibinfo{volume}{131}},
  \bibinfo{pages}{2766} (\bibinfo{year}{1963}).

\bibitem[{\citenamefont{Leichtle
  et~al.}(1996{\natexlab{a}})\citenamefont{Leichtle, Averbukh, and
  Schleich}}]{Leichtle:96a}
\bibinfo{author}{\bibfnamefont{C.}~\bibnamefont{Leichtle}},
  \bibinfo{author}{\bibfnamefont{I.}~\bibnamefont{Averbukh}}, \bibnamefont{and}
  \bibinfo{author}{\bibfnamefont{W.}~\bibnamefont{Schleich}},
  \bibinfo{journal}{Phys. Rev. Lett.} \textbf{\bibinfo{volume}{77}},
  \bibinfo{pages}{3999} (\bibinfo{year}{1996}{\natexlab{a}}).

\bibitem[{\citenamefont{Leichtle
  et~al.}(1996{\natexlab{b}})\citenamefont{Leichtle, Averbukh, and
  Schleich}}]{Leichtle:96b}
\bibinfo{author}{\bibfnamefont{C.}~\bibnamefont{Leichtle}},
  \bibinfo{author}{\bibfnamefont{I.}~\bibnamefont{Averbukh}}, \bibnamefont{and}
  \bibinfo{author}{\bibfnamefont{W.}~\bibnamefont{Schleich}},
  \bibinfo{journal}{Phys. Rev. A} \textbf{\bibinfo{volume}{54}},
  \bibinfo{pages}{5299} (\bibinfo{year}{1996}{\natexlab{b}}).

\bibitem[{\citenamefont{Barth and Lasser}(2009)}]{Barth2009}
\bibinfo{author}{\bibfnamefont{I.}~\bibnamefont{Barth}} \bibnamefont{and}
  \bibinfo{author}{\bibfnamefont{C.}~\bibnamefont{Lasser}},
  \bibinfo{journal}{J. Phys. B} \textbf{\bibinfo{volume}{42}},
  \bibinfo{pages}{235101} (\bibinfo{year}{2009}).

\bibitem[{\citenamefont{Gu\'{e}rin et~al.}(1997)\citenamefont{Gu\'{e}rin,
  Monti, Dupont, and Jauslin}}]{Guerin:97a}
\bibinfo{author}{\bibfnamefont{S.}~\bibnamefont{Gu\'{e}rin}},
  \bibinfo{author}{\bibfnamefont{F.}~\bibnamefont{Monti}},
  \bibinfo{author}{\bibfnamefont{J.-M.} \bibnamefont{Dupont}},
  \bibnamefont{and} \bibinfo{author}{\bibfnamefont{H.}~\bibnamefont{Jauslin}},
  \bibinfo{journal}{J. Phys. A} \textbf{\bibinfo{volume}{30}},
  \bibinfo{pages}{7193} (\bibinfo{year}{1997}).

\bibitem[{\citenamefont{Horenko et~al.}(2001)\citenamefont{Horenko, Schmidt,
  and Sch\"{u}tte}}]{BSchmidt:34}
\bibinfo{author}{\bibfnamefont{I.}~\bibnamefont{Horenko}},
  \bibinfo{author}{\bibfnamefont{B.}~\bibnamefont{Schmidt}}, \bibnamefont{and}
  \bibinfo{author}{\bibfnamefont{C.}~\bibnamefont{Sch\"{u}tte}},
  \bibinfo{journal}{J. Chem. Phys.} \textbf{\bibinfo{volume}{115}},
  \bibinfo{pages}{5733} (\bibinfo{year}{2001}).

\bibitem[{\citenamefont{Guerin and Jauslin}(2003)}]{Guerin:03a}
\bibinfo{author}{\bibfnamefont{S.}~\bibnamefont{Guerin}} \bibnamefont{and}
  \bibinfo{author}{\bibfnamefont{H.}~\bibnamefont{Jauslin}},
  \bibinfo{journal}{Adv. Chem. Phys.} \textbf{\bibinfo{volume}{125}},
  \bibinfo{pages}{147} (\bibinfo{year}{2003}).

\bibitem[{\citenamefont{Korolkov and Schmidt}(2004)}]{BSchmidt:41}
\bibinfo{author}{\bibfnamefont{M.~V.} \bibnamefont{Korolkov}} \bibnamefont{and}
  \bibinfo{author}{\bibfnamefont{B.}~\bibnamefont{Schmidt}},
  \bibinfo{journal}{Comp. Phys. Comm.} \textbf{\bibinfo{volume}{161}},
  \bibinfo{pages}{1} (\bibinfo{year}{2004}).

\bibitem[{\citenamefont{Breuer and Holthaus}(1989)}]{Breuer:89a}
\bibinfo{author}{\bibfnamefont{H.-P.} \bibnamefont{Breuer}} \bibnamefont{and}
  \bibinfo{author}{\bibfnamefont{M.}~\bibnamefont{Holthaus}},
  \bibinfo{journal}{Z. Phys. D} \textbf{\bibinfo{volume}{11}},
  \bibinfo{pages}{1} (\bibinfo{year}{1989}).

\bibitem[{\citenamefont{Drese and Holthaus}(1999)}]{Drese:99a}
\bibinfo{author}{\bibfnamefont{K.}~\bibnamefont{Drese}} \bibnamefont{and}
  \bibinfo{author}{\bibfnamefont{M.}~\bibnamefont{Holthaus}},
  \bibinfo{journal}{Eur. Phys. J. D} \textbf{\bibinfo{volume}{5}},
  \bibinfo{pages}{119} (\bibinfo{year}{1999}).

\bibitem[{\citenamefont{Worth and Cederbaum}(2004)}]{Worth:04a}
\bibinfo{author}{\bibfnamefont{G.~A.} \bibnamefont{Worth}} \bibnamefont{and}
  \bibinfo{author}{\bibfnamefont{L.~S.} \bibnamefont{Cederbaum}},
  \bibinfo{journal}{Ann. Rev. Phys. Chem.} \textbf{\bibinfo{volume}{55}},
  \bibinfo{pages}{127} (\bibinfo{year}{2004}).

\bibitem[{\citenamefont{{Fermanian Kammerer} and Lasser}(2003)}]{Lasser:05a}
\bibinfo{author}{\bibfnamefont{C.}~\bibnamefont{{Fermanian Kammerer}}}
  \bibnamefont{and} \bibinfo{author}{\bibfnamefont{C.}~\bibnamefont{Lasser}},
  \bibinfo{journal}{J. Math. Phys.} \textbf{\bibinfo{volume}{44}},
  \bibinfo{pages}{507} (\bibinfo{year}{2003}).

\bibitem[{\citenamefont{Truhlar and Mead}(2003)}]{Truhlar:03a}
\bibinfo{author}{\bibfnamefont{D.~G.} \bibnamefont{Truhlar}} \bibnamefont{and}
  \bibinfo{author}{\bibfnamefont{C.}~\bibnamefont{Mead}},
  \bibinfo{journal}{Phys. Rev. A} \textbf{\bibinfo{volume}{68}},
  \bibinfo{pages}{32501} (\bibinfo{year}{2003}).

\bibitem[{\citenamefont{Hagedorn and Joye}(2001)}]{Hagedorn2001}
\bibinfo{author}{\bibfnamefont{G.~A.} \bibnamefont{Hagedorn}} \bibnamefont{and}
  \bibinfo{author}{\bibfnamefont{A.}~\bibnamefont{Joye}},
  \bibinfo{journal}{Comm. Math. Phys.} \textbf{\bibinfo{volume}{223}},
  \bibinfo{pages}{583} (\bibinfo{year}{2001}).

\bibitem[{\citenamefont{Panati et~al.}(2007)\citenamefont{Panati, Spohn, and
  Teufel}}]{Panati2007}
\bibinfo{author}{\bibfnamefont{G.}~\bibnamefont{Panati}},
  \bibinfo{author}{\bibfnamefont{H.}~\bibnamefont{Spohn}}, \bibnamefont{and}
  \bibinfo{author}{\bibfnamefont{S.}~\bibnamefont{Teufel}},
  \bibinfo{journal}{Math. Mod. Num. Ana.} \textbf{\bibinfo{volume}{41}},
  \bibinfo{pages}{297} (\bibinfo{year}{2007}).

\bibitem[{\citenamefont{Horenko
  et~al.}(2002{\natexlab{a}})\citenamefont{Horenko, Salzmann, Schmidt, and
  Sch\"{u}tte}}]{BSchmidt:39}
\bibinfo{author}{\bibfnamefont{I.}~\bibnamefont{Horenko}},
  \bibinfo{author}{\bibfnamefont{C.}~\bibnamefont{Salzmann}},
  \bibinfo{author}{\bibfnamefont{B.}~\bibnamefont{Schmidt}}, \bibnamefont{and}
  \bibinfo{author}{\bibfnamefont{C.}~\bibnamefont{Sch\"{u}tte}},
  \bibinfo{journal}{J. Chem. Phys.} \textbf{\bibinfo{volume}{117}},
  \bibinfo{pages}{11075} (\bibinfo{year}{2002}{\natexlab{a}}).

\bibitem[{\citenamefont{Meyer}(1970)}]{Meyer:70a}
\bibinfo{author}{\bibfnamefont{R.}~\bibnamefont{Meyer}}, \bibinfo{journal}{J.
  Chem. Phys.} \textbf{\bibinfo{volume}{52}}, \bibinfo{pages}{2053}
  (\bibinfo{year}{1970}).

\bibitem[{\citenamefont{Marston and Balint-Kurti}(1989)}]{Marston:89a}
\bibinfo{author}{\bibfnamefont{C.~C.} \bibnamefont{Marston}} \bibnamefont{and}
  \bibinfo{author}{\bibfnamefont{G.~G.} \bibnamefont{Balint-Kurti}},
  \bibinfo{journal}{J. Chem. Phys.} \textbf{\bibinfo{volume}{91}},
  \bibinfo{pages}{3571} (\bibinfo{year}{1989}).

\bibitem[{\citenamefont{Webb and Hammes-Schiffer}(2000)}]{Webb:00a}
\bibinfo{author}{\bibfnamefont{S.~P.} \bibnamefont{Webb}} \bibnamefont{and}
  \bibinfo{author}{\bibfnamefont{S.}~\bibnamefont{Hammes-Schiffer}},
  \bibinfo{journal}{J. Chem. Phys.} \textbf{\bibinfo{volume}{113}},
  \bibinfo{pages}{5214} (\bibinfo{year}{2000}).

\bibitem[{\citenamefont{Manz et~al.}(2014)\citenamefont{Manz, Schild, Schmidt,
  and Yang}}]{BSchmidt:67}
\bibinfo{author}{\bibfnamefont{J.}~\bibnamefont{Manz}},
  \bibinfo{author}{\bibfnamefont{A.}~\bibnamefont{Schild}},
  \bibinfo{author}{\bibfnamefont{B.}~\bibnamefont{Schmidt}}, \bibnamefont{and}
  \bibinfo{author}{\bibfnamefont{Y.}~\bibnamefont{Yang}},
  \bibinfo{journal}{Chem. Phys.}  (\bibinfo{year}{2014}).

\bibitem[{\citenamefont{Gelman and Kosloff}(2003)}]{Gelman:03a}
\bibinfo{author}{\bibfnamefont{D.}~\bibnamefont{Gelman}} \bibnamefont{and}
  \bibinfo{author}{\bibfnamefont{R.}~\bibnamefont{Kosloff}},
  \bibinfo{journal}{Chem. Phys. Lett.} \textbf{\bibinfo{volume}{381}},
  \bibinfo{pages}{129} (\bibinfo{year}{2003}).

\bibitem[{\citenamefont{Korolkov
  et~al.}(1996{\natexlab{b}})\citenamefont{Korolkov, Manz, Paramonov, and
  Schmidt}}]{BSchmidt:17}
\bibinfo{author}{\bibfnamefont{M.~V.} \bibnamefont{Korolkov}},
  \bibinfo{author}{\bibfnamefont{J.}~\bibnamefont{Manz}},
  \bibinfo{author}{\bibfnamefont{G.~K.} \bibnamefont{Paramonov}},
  \bibnamefont{and} \bibinfo{author}{\bibfnamefont{B.}~\bibnamefont{Schmidt}},
  \bibinfo{journal}{Chem. Phys. Lett.} \textbf{\bibinfo{volume}{260}},
  \bibinfo{pages}{604} (\bibinfo{year}{1996}{\natexlab{b}}).

\bibitem[{\citenamefont{Hillery et~al.}(1984)\citenamefont{Hillery, O'Connell,
  Scully, and Wigner}}]{Hillery:84a}
\bibinfo{author}{\bibfnamefont{M.}~\bibnamefont{Hillery}},
  \bibinfo{author}{\bibfnamefont{R.}~\bibnamefont{O'Connell}},
  \bibinfo{author}{\bibfnamefont{M.}~\bibnamefont{Scully}}, \bibnamefont{and}
  \bibinfo{author}{\bibfnamefont{E.}~\bibnamefont{Wigner}},
  \bibinfo{journal}{Phys. Rep.} \textbf{\bibinfo{volume}{106}},
  \bibinfo{pages}{121} (\bibinfo{year}{1984}).

\bibitem[{\citenamefont{Buchholz et~al.}(2012)\citenamefont{Buchholz, Goletz,
  Gro{\ss}mann, Schmidt, Heyda, and Jungwirth}}]{BSchmidt:62}
\bibinfo{author}{\bibfnamefont{M.}~\bibnamefont{Buchholz}},
  \bibinfo{author}{\bibfnamefont{C.-M.} \bibnamefont{Goletz}},
  \bibinfo{author}{\bibfnamefont{F.}~\bibnamefont{Gro{\ss}mann}},
  \bibinfo{author}{\bibfnamefont{B.}~\bibnamefont{Schmidt}},
  \bibinfo{author}{\bibfnamefont{J.}~\bibnamefont{Heyda}}, \bibnamefont{and}
  \bibinfo{author}{\bibfnamefont{P.}~\bibnamefont{Jungwirth}},
  \bibinfo{journal}{J. Phys. Chem. A} \textbf{\bibinfo{volume}{116}},
  \bibinfo{pages}{11199} (\bibinfo{year}{2012}).

\bibitem[{\citenamefont{Meyer et~al.}(1990)\citenamefont{Meyer, Manthe, and
  Cederbaum}}]{Meyer:90a}
\bibinfo{author}{\bibfnamefont{H.-D.} \bibnamefont{Meyer}},
  \bibinfo{author}{\bibfnamefont{U.}~\bibnamefont{Manthe}}, \bibnamefont{and}
  \bibinfo{author}{\bibfnamefont{L.~S.} \bibnamefont{Cederbaum}},
  \bibinfo{journal}{Chem. Phys. Lett.} \textbf{\bibinfo{volume}{165}},
  \bibinfo{pages}{73} (\bibinfo{year}{1990}).

\bibitem[{\citenamefont{Manthe et~al.}(1992)\citenamefont{Manthe, Meyer, and
  Cederbaum}}]{Manthe:92a}
\bibinfo{author}{\bibfnamefont{U.}~\bibnamefont{Manthe}},
  \bibinfo{author}{\bibfnamefont{H.-D.} \bibnamefont{Meyer}}, \bibnamefont{and}
  \bibinfo{author}{\bibfnamefont{L.~S.} \bibnamefont{Cederbaum}},
  \bibinfo{journal}{J. Chem. Phys.} \textbf{\bibinfo{volume}{97}},
  \bibinfo{pages}{3199} (\bibinfo{year}{1992}).

\bibitem[{\citenamefont{Sch\"{a}fer-Bung
  et~al.}(2011)\citenamefont{Sch\"{a}fer-Bung, Hartmann, Schmidt, and
  Sch\"{u}tte}}]{BSchmidt:61}
\bibinfo{author}{\bibfnamefont{B.}~\bibnamefont{Sch\"{a}fer-Bung}},
  \bibinfo{author}{\bibfnamefont{C.}~\bibnamefont{Hartmann}},
  \bibinfo{author}{\bibfnamefont{B.}~\bibnamefont{Schmidt}}, \bibnamefont{and}
  \bibinfo{author}{\bibfnamefont{C.}~\bibnamefont{Sch\"{u}tte}},
  \bibinfo{journal}{J. Chem. Phys.} \textbf{\bibinfo{volume}{135}},
  \bibinfo{pages}{014112} (\bibinfo{year}{2011}).

\bibitem[{\citenamefont{Ohtsuki et~al.}(2004)\citenamefont{Ohtsuki, Turinici,
  and Rabitz}}]{Ohtsuki2004}
\bibinfo{author}{\bibfnamefont{Y.}~\bibnamefont{Ohtsuki}},
  \bibinfo{author}{\bibfnamefont{G.}~\bibnamefont{Turinici}}, \bibnamefont{and}
  \bibinfo{author}{\bibfnamefont{H.~A.} \bibnamefont{Rabitz}},
  \bibinfo{journal}{J. Chem. Phys.} \textbf{\bibinfo{volume}{120}},
  \bibinfo{pages}{5509} (\bibinfo{year}{2004}).

\bibitem[{\citenamefont{Horenko
  et~al.}(2002{\natexlab{b}})\citenamefont{Horenko, Schmidt, and
  Sch\"{u}tte}}]{BSchmidt:38}
\bibinfo{author}{\bibfnamefont{I.}~\bibnamefont{Horenko}},
  \bibinfo{author}{\bibfnamefont{B.}~\bibnamefont{Schmidt}}, \bibnamefont{and}
  \bibinfo{author}{\bibfnamefont{C.}~\bibnamefont{Sch\"{u}tte}},
  \bibinfo{journal}{J. Chem. Phys.} \textbf{\bibinfo{volume}{117}},
  \bibinfo{pages}{4643} (\bibinfo{year}{2002}{\natexlab{b}}).

\bibitem[{\citenamefont{Horenko et~al.}(2004)\citenamefont{Horenko, Weiser,
  Schmidt, and Sch\"{u}tte}}]{BSchmidt:43}
\bibinfo{author}{\bibfnamefont{I.}~\bibnamefont{Horenko}},
  \bibinfo{author}{\bibfnamefont{M.}~\bibnamefont{Weiser}},
  \bibinfo{author}{\bibfnamefont{B.}~\bibnamefont{Schmidt}}, \bibnamefont{and}
  \bibinfo{author}{\bibfnamefont{C.}~\bibnamefont{Sch\"{u}tte}},
  \bibinfo{journal}{J. Chem. Phys.} \textbf{\bibinfo{volume}{120}},
  \bibinfo{pages}{8913} (\bibinfo{year}{2004}).

\end{thebibliography}

\clearpage

\clearpage

\clearpage

\clearpage

\end{document}